\documentclass[%
 reprint,%
 amsmath,%
 amssymb,%
 aps,%
 floatfix,
 superscriptaddress]{revtex4-2}

\usepackage{xspace}
\usepackage{graphicx}
\graphicspath{{./figs}}
\usepackage{mathtools}
\usepackage{amsmath, amssymb}
\usepackage{dsfont}
\usepackage{bm}
\usepackage[dvipsnames]{xcolor}
\usepackage[colorlinks,%
            pdfusetitle,%
            bookmarksopen=false,%
            urlcolor=Blue,%
            citecolor=Blue,%
            linkcolor=Blue,%
            pdfencoding=auto,%
            psdextra]{hyperref}
\usepackage{physics}
\usepackage{upgreek}
\usepackage{bbm}
\usepackage[normalem]{ulem}
\usepackage{siunitx}
\usepackage{booktabs}

\newcommand{\ParaPlan}[1]{{\color{Violet}{#1}}}

\newcommand{\Ybferm}{\ensuremath{{}^\text{171}\text{Yb}}\xspace}
\newcommand{\Ptwo}{\ensuremath{{}^\text{3}\text{P}_2}\xspace}
\newcommand{\onePone}{\ensuremath{{}^\text{1}\text{P}_1}\xspace}
\newcommand{\Pone}{\ensuremath{{}^\text{3}\text{P}_1}\xspace}
\newcommand{\clock}{\ensuremath{{}^\text{3}\text{P}_0}\xspace}
\newcommand{\ground}{\ensuremath{{}^\text{1}\text{S}_0}\xspace}
\newcommand{\threeDone}{\ensuremath{{}^\text{3}\text{D}_1}\xspace}
\newcommand{\rydberg}{\ensuremath{ \ket{\nu\approx53.3,~L=0,~F=1/2,~m_F=-1/2~}}\xspace}

\newcommand{\gzero}{\ensuremath{\ket{g_0}}\xspace}

\newcommand{\mzero}{\ensuremath{\ket{m_0}}\xspace}
\newcommand{\mone}{\ensuremath{\ket{m_1}}\xspace}
\newcommand{\rzero}{\ensuremath{\ket{r_0}}\xspace}

\newcommand{\asciimathunit}[1]{\ensuremath{\,\mathrm{#1}}}
\newcommand{\um}{\ensuremath{\,\upmu \mathrm{m}}}
\newcommand{\uK}{\ensuremath{\,\upmu \mathrm{K}}}
\newcommand{\nm}{\asciimathunit{nm}}

\newcommand{\kHz}{\asciimathunit{kHz}}
\newcommand{\MHz}{\asciimathunit{MHz}}
\newcommand{\GHz}{\asciimathunit{GHz}}
\newcommand{\us}{\ensuremath{\,\upmu \mathrm{s}}}
\newcommand{\ms}{\asciimathunit{ms}}

\newcommand{\mW}{\asciimathunit{mW}}
\newcommand{\W}{\asciimathunit{W}}

\newcommand{\G}{\asciimathunit{G}}

\def \WavelengthUV{\ensuremath{302\nm}\xspace}
\def \WavelengthUVFundamentalTwo{\ensuremath{1582\nm}\xspace}
\def \WavelengthUVFundamentalOne{\ensuremath{977\nm}\xspace}
\def \WavelengthUVSFG{\ensuremath{604\nm}\xspace} 
\def \WavelengthTweezerOne{\ensuremath{532\nm}\xspace}
\def \WavelengthTweezerTwo{\ensuremath{767\nm}\xspace}
\def \WavelengthClockMagic{\ensuremath{759\nm}\xspace}

\def \WavelengthBlue{\ensuremath{399\nm}\xspace} 
\def \WavelengthGreen{\ensuremath{556\nm}\xspace} 
\def \WavelengthIonization{\ensuremath{369.5\nm}\xspace} 

\def \PowerUVSystem{\ensuremath{200\mW}\xspace}
\def \PowerUVSystemFiberAmplifiersMax{\ensuremath{10\W}\xspace}
\def \PowerUVSystemSFG{\ensuremath{1.5\W}\xspace}

\def \FreqMetaRabiFast{\ensuremath{2\pi\times 2.9 \MHz}\xspace} 
\def \FreqClockRabiMPP{\ensuremath{2\pi\times 90 \kHz}\xspace}

\def \FreqRydbergRabi{\ensuremath{2\pi\times 3 \MHz}\xspace}
\def \FreqRydbergRabiGHZ{\ensuremath{2\pi\times 3.0 \MHz}\xspace}
\def \FreqBlueImagingAltering{\ensuremath{3 \MHz}\xspace} 
\def \FreqRamseyDark{\ensuremath{3 \MHz}\xspace} 

\def \DetuningMetaSlow{\ensuremath{-2\pi\times 2 \GHz}\xspace}

\def \LifetimeRydberg{\ensuremath{46(2) \us}\xspace} 
\def \LifetimeRydbergSlow{\ensuremath{1.2(2) \ms}\xspace} 
\def \TimeRydbergPi{\ensuremath{0.15 \us}\xspace} 
\def \TimeMetaPi{\ensuremath{0.19 \us}\xspace} 
\def \TimeBlueImage{\ensuremath{15 \us}\xspace} 
\def \TimeBlueImageTwo{\ensuremath{17 \us}\xspace}
\def \TimeBlueImageOneTwo{\ensuremath{\text{15--17} \us}\xspace}
 
\def \TimeRamsey{\ensuremath{12.8(5)\us}\xspace}
\def \TimeTweezerRampTimeImaging{\ensuremath{0.5 \ms}\xspace}
\def \TimeProjectedThreeOutcomeMeasurement{\ensuremath{3}~\ms\xspace}

\def \FidelityBellMetaLossdetected{\ensuremath{98.6(3)\%}\xspace} 
\def \FidelityBellMetaLossdetectedPop{\ensuremath{98.7(2)\%}\xspace} 
\def \FidelityBellMetaLossdetectedParity{\ensuremath{98.4(6)\%}\xspace}

\def \FidelityBellMetaLossdetectedMcorr{\ensuremath{99.0(3)\%}\xspace} 
\def \FidelityBellMetaLossdetectedMnoflip{\ensuremath{98.7(3)\%}\xspace} 

\def \FidelityBellClockLossdetected{\ensuremath{93.8(6)\%}\xspace} 
\def \FidelityBellClockLossdetectedPop{\ensuremath{98.6(2)\%}\xspace} 
\def \FidelityBellClockLossdetectedParity{\ensuremath{89.1(1)\%}\xspace} 

\def \FidelityPXPBellMetaLossdetected{\ensuremath{96.3(8)\%}\xspace} 
\def \FidelityPXPBellMetaLossdetectedPop{\ensuremath{96.8(3)\%}\xspace} 
\def \FidelityPXPBellMetaLossdetectedParity{\ensuremath{95.8(1.5)\%}\xspace} 

\def \FidelityPXPBellClockLossdetected{\ensuremath{92.4(9)\%}\xspace} 
\def \FidelityPXPBellClockLossdetectedPop{\ensuremath{96.4(3)\%}\xspace} 
\def \FidelityPXPBellClockLossdetectedParity{\ensuremath{88.3(1.8)\%}\xspace} 

\def \FidelitygRBNoEchoLossdetected{\ensuremath{99.78(4)\%}\xspace}
\def \FidelitygRBNoEchoErasure{\ensuremath{99.51(6)\%}\xspace}
\def \FidelitygRBNoEchoRaw{\ensuremath{99.37(7)\%}\xspace}
\def \FidelitygRBEchoLossdetected{\ensuremath{99.75(4)\%}\xspace}

\def \InFidelityCZTheory{\ensuremath{0.35 \%}\xspace}
\def \InFidelityCZTheoryLossdetected{\ensuremath{0.021\%}\xspace}
\def \InFidelityCZTheoryExpDescrepancy{\ensuremath{0.2\%}\xspace}
\def \AverageXrot{\ensuremath{1.6}\xspace}

\def \FidelityGHZTwentyatoms{\ensuremath{51(4)\%}\xspace}
\def \FidelityGHZTwentyatomsZTwo{\ensuremath{78(3)\%}\xspace}

\def \FidelityZTwoPopPrepErasure{\ensuremath{53(3)\%}\xspace}
\def \FidelityZTwoPopDecayErasyre{\ensuremath{57(4)\%}\xspace}
\def \FidelityZTwoPopLossdetected{\ensuremath{78(3)\%}\xspace}

\def \FidelityRydbergDetectionEfficiencyCoarse{\ensuremath{>90\%}\xspace}

\def \FidelityRydbergDetectionEfficiency{\ensuremath{93(1)\%}\xspace}
\def \FidelityRydbergDetectionEfficiencySingle{\ensuremath{96(1)\%}\xspace}
\def \FidelityRydbergDetectionEfficiencySingleDecayComparison{\ensuremath{96.1(7) \%}\xspace}

\def \InfidelityBlueImagingErasure{\ensuremath{<0.5 \%}\xspace}
\def \InfidelityBlueImagingSpin{\ensuremath{<0.2 \%}\xspace}

\def \EfficiencyLoadingTypical{\ensuremath{\sim90\%}\xspace}
\def \EfficiencySortingTypical{\ensuremath{>50\%}\xspace}

\def \InfidelitySpinMisidentification{\ensuremath{0.44(6)\%}\xspace}
\def \InfidelitySpinMisidentificationSpinLoss{\ensuremath{3.6(1)\%}\xspace}
\def \InfidelityRydbergProjectionDecay{\ensuremath{0.5\%}\xspace}

\def \NonHInitPop{\ensuremath{\{16(1)\%, 53(2)\%, 84(1)\%, 99.7(3)\%\}}\xspace}

\def \BranchThreeSoneToThreePzero{\ensuremath{15\%}\xspace}
\def \RydbergBranchthreePtwoRatio{\ensuremath{48(6)\%}\xspace}
\def \RydbergBranchPop{\ensuremath{\{42(2)\%, 20(2)\%, 2.7(5)\%, 4(1)\%\}}\xspace}
\def \RydbergBranchOtherPop{\ensuremath{31(3)\%}\xspace}
\def \LengthscaleGateDistance{\ensuremath{2.4\um}\xspace}
 
\def \LengthscaleRamsey{\ensuremath{12\um}\xspace}

\def \NumbersRabiClockFlip{\ensuremath{>60}\xspace}
\def \NumbersRabiBlockadedFlip{\ensuremath{72(6)}\xspace}
\def \NumbersRabiSingleFlip{\ensuremath{70(10)}\xspace} 
\def \NumbersNbarRaman{0.05\xspace} 
\def \NumbersTempMonteCarlo{0.28(4)\uK} 
\def \NumbersSortedArraySize{\ensuremath{40}\xspace}
\def \NumbersSortedArrayFraction{\ensuremath{>60\%}\xspace}
\def \NumbersSortedArraySizeMax{\ensuremath{50}\xspace}
\def \NumbersSortedArrayFractionMax{\ensuremath{>20\%}\xspace}
\def \NumbersArrayMax{\ensuremath{64}\xspace}
\def \NumbersBField{\ensuremath{8.2\G}\xspace} 
\def \NumbersEFieldGradient{\ensuremath{0.62 \kHz/ \um}\xspace} 
\def \NumbersRydbergRabiContrastLossdetected{\ensuremath{99.6(3)\%}\xspace}
\def \FreqQubitSplit{\ensuremath{2\pi\times 9.5 \kHz}\xspace}
\def \DiagonalEnergy{\ensuremath{\sim 2\pi\times 1 \MHz}\xspace}
\def \NumberProjectionDetuning{\ensuremath{ 2\pi\times 1.0 \MHz}\xspace}
\def \NumbersFinesseULERydberg{\ensuremath{\sim2\times10^4}\xspace}

\def \NumbersShallowTweezerDepth{\ensuremath{250 \kHz}\xspace}
\def \NumbersDeepTweezerDepth{\ensuremath{9.6 \MHz}\xspace}

\def \NumberIonPower{\ensuremath{1 \mW}\xspace}
\def \NumberIonWasit{\ensuremath{40\um}\xspace}
\def \NumberIonDecay{\ensuremath{0.03\us}\xspace}
\def \NumberIonDecayErrLim{\ensuremath{0.1\%}\xspace}

\def \NumbersRydbergLatticeDeformation{\ensuremath{3\%}\xspace}
\def \NumbersgRBInst{\ensuremath{40}\xspace}

\renewcommand{\figurename}{Fig.}

\begin{document}

\title{High-fidelity entanglement and coherent multi-qubit mapping in an atom array}

\author{Aruku Senoo}
\affiliation{%
JILA, University of Colorado and National Institute of Standards and Technology,
and Department of Physics, University of Colorado, Boulder, Colorado 80309, USA
}%

\author{Alexander Baumgärtner}
\affiliation{%
JILA, University of Colorado and National Institute of Standards and Technology,
and Department of Physics, University of Colorado, Boulder, Colorado 80309, USA
}%

\author{Joanna W. Lis}
\affiliation{%
JILA, University of Colorado and National Institute of Standards and Technology,
and Department of Physics, University of Colorado, Boulder, Colorado 80309, USA
}%

\author{Gaurav M. Vaidya}
\affiliation{%
JILA, University of Colorado and National Institute of Standards and Technology,
and Department of Physics, University of Colorado, Boulder, Colorado 80309, USA
}%

\author{Zhongda Zeng}
\affiliation{Institute for Theoretical Physics, University of Innsbruck, Innsbruck A-6020, Austria
}%
\affiliation{Institute for Quantum Optics and Quantum Information, Austrian Academy of Sciences, Innsbruck A-6020, Austria}%

\author{Giuliano Giudici}
\affiliation{Institute for Theoretical Physics, University of Innsbruck, Innsbruck A-6020, Austria
}%
\affiliation{Institute for Quantum Optics and Quantum Information, Austrian Academy of Sciences, Innsbruck A-6020, Austria}%

\author{Hannes Pichler}
\affiliation{Institute for Theoretical Physics, University of Innsbruck, Innsbruck A-6020, Austria
}%
\affiliation{Institute for Quantum Optics and Quantum Information, Austrian Academy of Sciences, Innsbruck A-6020, Austria}%

\author{Adam M. Kaufman}
\email[e-mail:$\,$]{adam.kaufman@colorado.edu}
\affiliation{%
JILA, University of Colorado and National Institute of Standards and Technology,
and Department of Physics, University of Colorado, Boulder, Colorado 80309, USA
}%
\date{\today}

\begin{abstract}

Neutral atoms in optical tweezer arrays possess broad applicability for quantum information science, in computing~\cite{saffman2016quantum,bluvstein2024logical,graham2022multi}, simulation~\cite{browaeys2020many,semeghini2021probing,shaw2024benchmarking}, and metrology~\cite{norcia2019seconds,madjarov2019atomic,young2020half}. Among atomic species, Ytterbium-171 is unique as it hosts multiple qubits, each of which is impactful for these distinct applications~\cite{jenkins2022ytterbium,ma2022universal,ma2023high,lis2023midcircuit,muniz2025high,ludlow2015optical}. Consequently, this atom is an ideal candidate to bridge multiple disciplines, which, more broadly, has been an increasingly effective strategy within the field of quantum science~\cite{pezze2018quantum,andersen2025thermalization,bluvstein2022quantum,huang2022quantum,finkelstein2024universal,eckner2023realizing,marciniak2022optimal}. Realizing the full potential of this synergy requires high-fidelity generation and transfer of many-particle entanglement between these distinct qubit degrees of freedom, and thus between these distinct applications.  Here we demonstrate the creation and coherent mapping of entangled quantum states across multiple qubits in Ytterbium-171 tweezer arrays. We map entangled states onto the optical clock qubit~\cite{pedrozo2020entanglement,schine2022long,cao2024multi, finkelstein2024universal} from the nuclear spin qubit~\cite{jenkins2022ytterbium, ma2022universal, barnes2021assembly} or the Rydberg qubit~\cite{levine2018high,madjarov2020high}. 
We coherently transfer up to 20 atoms of a $Z_2$-ordered Greenberger-Horne-Zeilinger (GHZ) state~\cite{omran2019generation} from the interacting Rydberg manifold to the metastable nuclear spin manifold. The many-body state is generated via a novel disorder-robust pulse in a two-dimensional ladder geometry. We further find that clock-qubit-based spin detection applied to Rydberg and nuclear spin qubits facilitates atom-loss-detectable qubit measurements and \FidelityRydbergDetectionEfficiencyCoarse Rydberg decay detection~\cite{ma2023high,lis2023midcircuit,muniz2025high}. This enables mid-circuit and delayed erasure detection, yielding an error-detected two-qubit gate fidelity of \FidelitygRBNoEchoLossdetected~\cite{radnaev2025universal,muniz2025high} in the metastable qubits. This error detection also enables Rydberg qubit evolution with an effective lifetime of \LifetimeRydbergSlow, enhancing the fidelity of the observed many-body dynamics.
These results establish a versatile architecture that advances multiple fields of quantum information science while also establishing bridges between them.

\end{abstract}

\maketitle

\begin{figure*}[t]
    \centering
    \includegraphics[width=\textwidth]{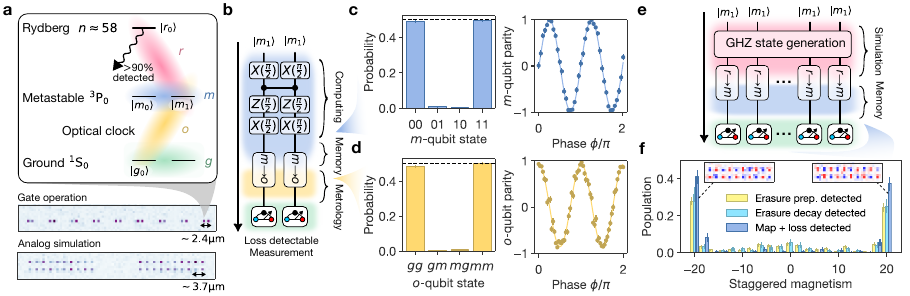}
    \caption{\label{fig1}%
    \textbf{Architecture based on multi-level qubit mapping.}
    \textbf{a,}~ Schematic of the \Ybferm atomic structure (top panel). Coherent spin-1/2 nuclear spin qubit in both the ground state ($g$-qubit, green) and the metastable state ($m$-qubit, blue). An optical transition connects the ground and metastable states, defining the clock qubit ($o$-qubit, yellow). The $m_F=+1/2$ metastable state can be excited to a Rydberg level via a single photon transition at \WavelengthUV ($r$-qubit, red). 
    In this system, \FidelityRydbergDetectionEfficiencyCoarse of Rydberg decay events are detected as atom loss. The bottom panel shows single-shot images after rearrangement for experiments involving two-qubit gates and experiments using analog many-body Hamiltonian evolutions.  \textbf{b,}~ Bell state generation and mapping sequence. The Bell state generated by a high-fidelity two-qubit gate in the $m$-qubit manifold is subsequently mapped onto the $o$-qubit and detected by loss-detectable spin-measurements represented by a box with three dots. \textbf{c,}~ Measured $m$-qubit Bell state population (left) and the parity oscillation (right) with loss detection. \textbf{d,}~ Analogous measurement for $o$-qubit. \textbf{e,}~ The sequence of the GHZ-state mapping from $r$-qubit to $m$-qubit. The GHZ state is generated by an adiabatic many-body Hamiltonian sweep in the $r$-qubit and mapped to $m$-qubit. \textbf{f,}~ Measured $Z_2$-order of the GHZ state. The staggered magnetism is defined as a sum of the magnetism with flipped signs for the nearest neighbor atoms (Methods). Three spin detection methods are compared. Yellow (left) shows the data of direct $r$-qubit spin detection with erasure detection to suppress preparation errors (Methods). Light blue (center) includes additional error detection through erasures revealed after the sequence, which identifies Rydberg-state decay events. Finally, dark blue (right) applies qubit mapping and loss detection, as shown in e. The observed $Z_2$ population is \FidelityZTwoPopPrepErasure, \FidelityZTwoPopDecayErasyre, and \FidelityZTwoPopLossdetected respectively, showing two-fold improvement of population error in the loss detected case. (Inset) Single-shot loss-detected measurement result; images for two spins are combined with blue (red) color for \mzero (\mone). 
    }
\end{figure*}

Quantum technologies are advancing along parallel tracks, with breakthroughs in computation~\cite{kjaergaard2020superconducting,bruzewicz2019trapped,henriet2020quantum,ai2024quantum,bluvstein2024logical}, simulation~\cite{georgescu2014quantum,daley2022practical}, and metrology~\cite{ludlow2015optical,pezze2018quantum}. Typically, each realization is driven by a specific qubit --- a fundamental unit of quantum information --- optimized to a particular task. For instance, a qubit optimized for sensing is unlikely to be ideal for quantum computing, as the sensitivity of the former compromises the robustness of the latter. However, combining such distinct functionalities unlocks a variety of new possibilities. Digital quantum gates can enhance access to observables previously challenging to measure for analog quantum simulations~\cite{bluvstein2022quantum,lamata2018digital,andersen2025thermalization}; non-classical states generated through computation or simulation can be mapped onto ultranarrow optical transitions for quantum-enhanced metrology~\cite{pedrozo2020entanglement,pezze2018quantum,kaubruegger2023optimal, marciniak2022optimal}; and, computing-inspired error mitigation methods can be translated to quantum simulation and metrology~\cite{zhou2018achieving,kielinski2024ghz, evered2025probing}.

In this context, favorable features of optical-tweezer-trapped neutral atoms --- including isolation, scalability, and precision control --- have found utility in multiple directions in quantum science, making it a promising platform to explore such synergies. These features, along with the non-local connectivity afforded by mobile optical tweezers, offer a promising framework for quantum computing~\cite{saffman2016quantum, bluvstein2024logical, reichardt2024logical,chinnarasu2025variational}, while their flexible geometry and tunable interactions are enabling for quantum simulation~\cite{browaeys2020many,semeghini2021probing,shaw2024benchmarking}. Additionally, the ability to isolate single atoms close to the motional ground state in a collision-free manner makes them well suited for metrology~\cite{norcia2019seconds,madjarov2019atomic,young2020half}. 

Among atomic species, Ytterbium-171 (\Ybferm) is notable for its unique atomic level structure (Fig.~\ref{fig1}a), which supports multiple qubits~\cite{lis2023midcircuit}. For quantum computation and memory applications, the spin $1/2$-nuclear spin qubit, hosted in both the ground ($g$) and metastable ($m$) electronic states, is particularly attractive due to its long coherence times of up to 10 seconds~\cite{lis2023midcircuit}, arising from the absence of hyperfine interactions with orbital electrons~\cite{ma2022universal, jenkins2022ytterbium,lis2023midcircuit,norcia2023midcircuit}. The transition connecting the ground and metastable state (the optical qubit, $o$) is the basis of world-leading optical atomic clocks~\cite{ludlow2015optical}. The single-photon transition between the metastable clock state and $^3\mathrm{S}_1$ Rydberg series ($r$-qubits) enables high-fidelity entanglement generation ~\cite{madjarov2020high,ma2023high,eckner2023realizing,cao2024multi,scholl2023erasure,tsai2025benchmarking,peper2025spectroscopy}. And, most importantly, these different qubits can cooperate effectively: Rydberg interactions can be used to engineer entanglement in any of these qubits via digital or analog protocols, while the clock qubit allows loss-detectable spin measurements, which can be leveraged for enhanced fidelity~\cite{lis2023midcircuit}. 
Despite the broad potential, so far, combining all of these capabilities has yet to be achieved. 

In this work, we demonstrate the full realization of this integration. Specifically, we create clock-qubit entangled states, generated either by two-qubit gates on the metastable state (Fig.~\ref{fig1}b) or analog Hamiltonian evolution. Extending this capability to large entangled states, we map $Z_2$-ordered Greenberger-Horne-Zeilinger (GHZ) states of up to 20 atoms onto long-lived nuclear spin qubits (Fig.~\ref{fig1}e).  
Notably, our near-adiabatic preparation of the GHZ state in a ladder geometry exploits a novel disorder-robust scheme~\cite{zeng2025adiabatic} discovered via quantum optimal control techniques~\cite{Glaser2015}.  In all of these demonstrations, we exploit clock-shelving-based loss detection to each qubit type --- interacting Rydberg qubits, nuclear-spin qubits, and clock qubits ---  leading to detection of \FidelityRydbergDetectionEfficiencyCoarse of Rydberg decay errors during analog Hamiltonian evolution with effective Rydberg lifetime of \LifetimeRydbergSlow~(Fig.~\ref{fig1}f) and an error-detected two-qubit gate fidelity of \FidelitygRBNoEchoLossdetected with corresponding high-fidelity Bell-state generation (Fig.~\ref{fig1}c). These results establish a versatile multi-qubit platform, providing a compelling new strategy to seamlessly integrate quantum computing, simulation, and metrology.

\subsection*{Multi-level qubit mapping and loss detection}

\begin{figure*}[t]
    \centering
    \includegraphics[width=\textwidth]{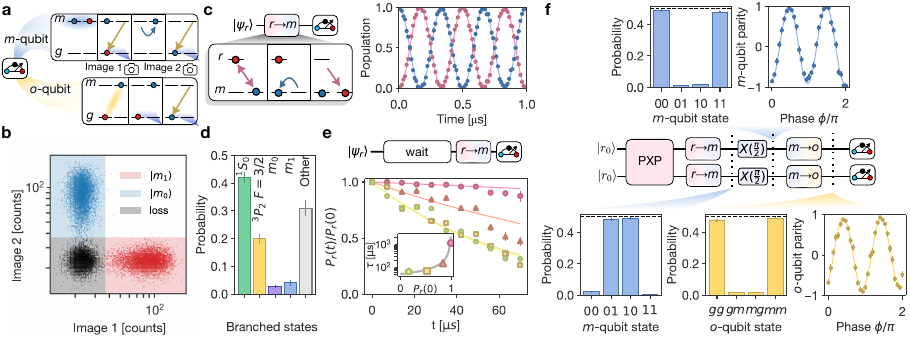}
    \caption{\label{fig2}%
    \textbf{Qubit mapping and loss-detectable spin measurement.}
    \textbf{a,}~ Sequence of the loss-detectable spin measurements for the $m$-qubit (top) and the $o$-qubit (bottom). A clock $\pi$-pulse is used to selectively de-excite one of the spin components. Each spin component is destructively imaged via fast fluorescence imaging (see Methods). \textbf{b,}~ Distribution of the camera counts of the spin measurement. We apply the measurement to distinguish the three outcomes \mone, \mzero, and atom loss, for the case of preparing each of these states (red, blue, and black dots). \textbf{c,}~ Loss-detectable spin measurement of $r$-qubit. The left schematic shows the sequence of mapping the $r$-qubit onto the $m$-qubit. (Right) Rabi oscillations of the Rydberg state observed by this spin-measurement method with loss detection. Both \rzero population (red) and \mone population (blue) are observed simultaneously. \textbf{d,}~ Measured branching ratio of the decay for \rzero Rydberg state (see Extended Data Fig.~\ref{SMfig4_5}). Decays are registered as atom loss unless the atom decays back into the qubit manifold in use. \textbf{e,}~ Observation of population decay dynamics consistent with the non-Hermitian state evolution via qubit mapping and loss detection. The loss-detected probability of Rydberg state, $P_r(t)$ is plotted, where each marker represents different initial Rydberg populations, \NonHInitPop. The inset compares the fitted decay time constant and the curve derived from non-Hermitian dynamics without free parameters (see Methods). We use $P_r(t)=P_r(0)e^{-t/\tau}$ for the fitting function. \textbf{f,}~ Proof-of-concept demonstration combining analog quantum simulation and digital quantum operations to generate a metrologically valuable state. After evolving two atoms under the PXP-Hamiltonian, a Bell state in the Rydberg qubit is generated and mapped onto the $m$-qubit (lower left). A $\pi/2$-pulse on the $m$-qubit then converts the $(\ket{01}+\ket{10})/\sqrt{2}$ Bell state into a $(\ket{00}+\ket{11})/\sqrt{2}$ Bell state (top), which is subsequently mapped onto the clock $o$-qubit (lower right). 
    }
\end{figure*}

We manipulate \Ybferm trapped in optical tweezer arrays in a two-dimensional ladder arrangement. The single atom arrays are prepared by combining  \EfficiencyLoadingTypical enhanced loading and rearrangement~\cite{jenkins2022ytterbium, browaeys2020many}~(Extended Data Fig.~\ref{SMfig1} and Methods). For loading and rearrangement, we use \WavelengthTweezerOne tweezer arrays. Atoms are then transferred to \WavelengthTweezerTwo tweezers close to the clock-magic wavelength (Extended Data Fig.~\ref{SMfig1} and Methods).
In each experimental run, we initialize the atoms in the radial motional ground state and coherently excite them to the metastable state via the optical clock transition, while preserving their motional level~\cite{lis2023midcircuit}.

Our study focuses on three distinct qubit types (Fig.~\ref{fig1}a), out of 4 atomic levels: the $m_F=-1/2$ hyperfine state of the ground state \ground (\gzero), the $m_F=\mp1/2$ hyperfine states of the metastable \clock (\mzero and \mone), and the \rydberg ~Rydberg state with $n\approx58$ (\rzero) identified in~\cite{peper2025spectroscopy}.
The $o$-qubit widely employed in optical clock metrology consists of \gzero and \mone ~\cite{ludlow2015optical}. We implement global single-qubit $X$-rotations for this qubit with a typical Rabi frequency of \FreqClockRabiMPP, while virtual $Z$-rotations are controlled with the optical phase. The $m$-qubit is encoded in the nuclear spin states \mzero and \mone~\cite{lis2023midcircuit,ma2022universal}, with $X$-rotations performed using optical Raman rotations~\cite{lis2023midcircuit}. To minimize the scattering error during the single-qubit operations, we employ phase tracking for $Z$-rotations where our qubit splitting is \FreqQubitSplit in the typical magnetic field of \NumbersBField. Finally, the $r$-qubit is defined between \mone and \rzero. Using a \WavelengthUV ultraviolet (UV) laser, we drive the qubit with a Rabi frequency of around \FreqRydbergRabi via a single-photon excitation.

Fig.~\ref{fig1} visualizes the mapping of an entangled state. We first show the transfer of the entangled Bell state from the $m$-qubit to the $o$-qubit (see Fig.~\ref{fig1}b). 
Using a high-fidelity two-qubit gate (detailed later), we generate the $(\ket{00}-i\ket{11})/\sqrt{2}$ Bell state in the $m$-qubit manifold. The loss-detected fidelity of this Bell state is \FidelityBellMetaLossdetected based on the average of the Bell state population, \FidelityBellMetaLossdetectedPop, and amplitude of parity oscillations, \FidelityBellMetaLossdetectedParity(Fig.~\ref{fig1}c); we estimate \FidelityBellMetaLossdetectedMcorr Bell state fidelity after correcting for measurement errors (Methods). To transfer the entangled state to the $o$-qubit, we apply two consecutive $\pi$-pulses first to the $o$-, then the $m$-manifold. The loss-detected fidelity of the transferred Bell state results in \FidelityBellClockLossdetected~with a population of \FidelityBellClockLossdetectedPop and parity oscillation amplitude of \FidelityBellClockLossdetectedParity~(Fig.~\ref{fig1}d). During the mapping to the $o$-qubit, tweezers are turned off to avoid differential light shift from the non-magic wavelength tweezer.

We exploit loss-detectable spin measurements of the metastable qubit for error detection, which have recently been developed in several qubit architectures in neutral atom platforms~\cite{lis2023midcircuit,norcia2023midcircuit,deist2022mid,radnaev2025universal,huie2023repetitive,hu2025site,evered2025probing}. 
Here, we employ the selectivity of the polarization of the clock beam for hiding one of the states in the metastable manifold~\cite{lis2023midcircuit}.
The sequential imaging of both atomic spin states allows measurement of both spin states as well as atom loss. While previous demonstrations concentrate on measuring $g$-qubit loss-detectably~\cite{lis2023midcircuit,norcia2023midcircuit, huie2023repetitive}, we expand this to multiple qubits applying the mapping. Fig.~\ref{fig2}a illustrates the detailed procedure for the spin measurement in both $m$-qubit and $o$-qubit. The distribution of the camera counts in Fig.~\ref{fig2}b compares the three situations where atoms are prepared in \mzero or \mone and no atom is prepared. From this, the average spin-detection infidelity is deduced as \InfidelitySpinMisidentification, while the False-Positive rate of the loss detection is \InfidelitySpinMisidentificationSpinLoss due to the extra loss induced during the measurement process~(see Extended Data Table.~\ref{table1} and Methods).

The loss-detectable spin measurement can be extended to the $r$-qubit by mapping the $r$-qubit spin state onto the $m$-qubit. Fig.~\ref{fig2}c illustrates this procedure, which consists of a $\pi$-pulse on the $m$-qubit followed by a $\pi$-pulse on the $r$-qubit before doing the $m$-qubit spin measurement. To mitigate the effect of dephasing the $r$-qubit during this process, the $m$-qubit rotation must be much faster than the coherence time of the Rydberg state. In this experiment, the $T_2^*$ coherence time is \TimeRamsey and the Rydberg decay time constant is \LifetimeRydberg (see Extended Data Fig.~\ref{SMfig4} and Methods).
By using the smaller intermediate state detuning of our two operation modes, we achieve a \TimeMetaPi $\pi$-pulse duration of $m$-qubit $X$ rotation, a comparable speed to the $r$-qubit $\pi$-pulse of \TimeRydbergPi (see Extended Data Fig.~\ref{SMfig3} and Methods). This suppresses the estimated Rydberg decayed population during the process to below \InfidelityRydbergProjectionDecay per atom. 
The right panel of Fig.~\ref{fig2}c shows Rabi oscillations on the $r$-qubit measured with this scheme, where the amplitude contrast is observed to be as high as \NumbersRydbergRabiContrastLossdetected, due to the mitigation of preparation errors and high-fidelity control of the Rydberg state (see Extended Data Fig.~\ref{SMfig4} and Methods).

The loss-detectable $r$-qubit spin measurement is naturally applied to detect the decay of the Rydberg state, as most decay pathways lead to loss. The fidelity of the decay detection is governed by the branching ratio of the Rydberg state, presented in Fig.~\ref{fig2}d. 
For experiments based on the $r$-qubit, since we use \rzero and \mone, our Rydberg decay detection fidelity is \FidelityRydbergDetectionEfficiencySingle after the application of the resonant scattering beam to remove any ground state population. Also, the fact that the \Ptwo metastable state is highly anti-trapped in our \WavelengthTweezerTwo tweezers prevents them from decaying back to the ground state manifold during the spin read-out. For the experiment involving both $m$-qubit and $r$-qubit, such as the two-qubit gate experiments, the decay detection fidelity reduces to \FidelityRydbergDetectionEfficiency, accounting for the combined decay rates to \mzero and \mone. However, because this loss-detection-based method is effective to additional populations that do not decay (or get repumped) to the ground state, it leads to a significant improvement in the total fraction of detected erasures relative to previous experiments~\cite{scholl2023erasure,ma2023high, muniz2025high}, at the price of delayed information arrival until qubit detection is performed.

The high fraction of detected decay events can reveal the non-Hermitian state evolution of a decaying quantum system~\cite{molmer1993monte}. As a simple example, we observe the effective population evolution of superposition between a decaying $\ket{r_0}$ state and a non-decaying $\ket{m_1}$ state. By preparing a coherent superposition $\ket{\psi_r}=\sqrt{1-|a|^2}\ket{m_1}+a\ket{r_0}$, which has a Rydberg fraction of $|a|^2$, we investigate how the Rydberg state population, $P_r(t)$, evolves in experiments where Rydberg decay is not detected. When the initial population is entirely in the Rydberg state, the effective population decay is suppressed. However, as the portion of \mone increases, we detect an accelerated decay rate of this population, illustrating that even without observing the decay, the populations evolve (Fig.~\ref{fig2}e). We compare our observation with the expected behavior of the non-Hermitian evolution, which shows good agreement (Fig.~\ref{fig2}e inset and Methods). The observed slow decay time constant of \LifetimeRydbergSlow for the pure Rydberg atom case ($P_r(0)=1$) indicates the fidelity of the decay detection is \FidelityRydbergDetectionEfficiencySingleDecayComparison, consistent with the estimate from the branching ratios.

To further illustrate the flexibility of the multi-qubit mapping, we demonstrate the transfer of the $r$-qubit Bell state to the $m$-qubit and then to the $o$-qubit, conceptually illustrating the combination of quantum simulation, quantum computation, and quantum metrology (Fig.~\ref{fig2}f). In the highly Rydberg-blockaded regime, two atoms evolve according to the ideal PXP model Hamiltonian, generating the $(\ket{mr}+\ket{rm})/\sqrt{2}$ Bell state in the $r$-qubit~\cite{browaeys2020many}. We halt the evolution at this point, and map the $r$-qubit state to the $m$-qubit. By applying a X $\pi/2$-pulse to the $(\ket{01}+\ket{10})/\sqrt{2}$ Bell state in the $m$-qubit, we generate the metrologically-useful $(\ket{00}+\ket{11})/\sqrt{2}$ Bell state, where the loss-detected fidelity is measured to be \FidelityPXPBellMetaLossdetected with population of \FidelityPXPBellMetaLossdetectedPop and parity oscillation amplitude of \FidelityPXPBellMetaLossdetectedParity. Finally, we map the $m$-qubit Bell state to the $o$-qubit, reaching \FidelityPXPBellClockLossdetected fidelity with \FidelityPXPBellClockLossdetectedPop population and \FidelityPXPBellClockLossdetectedParity parity oscillation amplitude.

\subsection*{Rydberg-mediated two-qubit gates with loss detection}

\begin{figure}[t]
    \centering
    \includegraphics[width=\columnwidth]{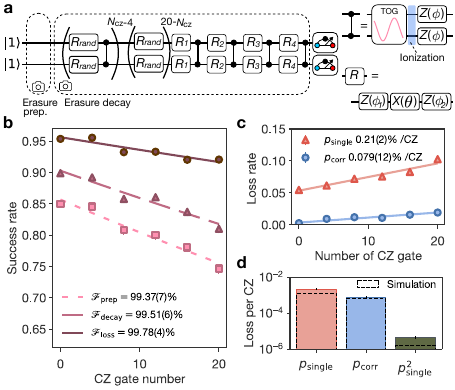}
    \caption{\label{fig3}%
    \textbf{High-fidelity two-qubit gate enhanced by loss detection.}
     \textbf{a,}~ Global randomized benchmarking (gRB) sequence. To compare to the loss detection, ground state erasure detection is implemented. Preparation errors in the metastable state are detected by observing the ground state population at the beginning of the gRB sequence, while some of the qubit leakage errors are continuously monitored by observing the ground state population during the sequence. The erasure detection beam is off only during the two-qubit gate operations. \textbf{b,}~ gRB results. We observe an increase in the gate fidelity by partially detecting the decayed population by erasure decay detection (triangle) compared to the case only with preparation erasure detection (square). With loss detection, the fidelity increases to \FidelitygRBNoEchoLossdetected (circle).  \textbf{c,}~  Atom loss in the gRB experiment after preparation erasure detection. Here, the loss includes auto-ionized population due to the remaining population in the Rydberg state after a gate.  \textbf{d,}~ Histogram of measured loss per CZ gate. The dashed line shows the simulation result (Methods). 
    }
\end{figure}

Applying the loss- and Rydberg-decay-detection, we demonstrate high-fidelity two-qubit gates~\cite{muniz2025high, radnaev2025universal} and studies of unique errors arising during the Rydberg-blockade dynamics underlying the gate. We employ a time-optimal gate (TOG) protocol ~\cite{jandura2022time, evered2023high}, sinusoidally modulating the optical phase of the driving laser. To evaluate the two-qubit gate fidelity separate from SPAM errors and single-qubit errors, a global randomized benchmarking (gRB) is implemented~\cite{evered2023high,ma2023high,tsai2025benchmarking,radnaev2025universal,muniz2025high,peper2025spectroscopy}. We alternate randomly chosen single-qubit Clifford gates with two-qubit gate operations, and the final four single-qubit gates, $R_1,R_2,R_3,R_4$, are chosen such that both qubits return to their initial state.
Importantly, we also apply an ionization pulse after each CZ gate to convert residual Rydberg population after the gate to (detectable) loss.

With this approach, we extract a raw gate fidelity of \FidelitygRBNoEchoRaw (see Methods). While detection of Rydberg decay via the ground-state population improves the fidelity to \FidelitygRBNoEchoErasure, loss detection further improves the fidelity, observing \FidelitygRBNoEchoLossdetected two-qubit gate fidelity. The relative improvement in fidelity using erasure decay detection compared to loss detection implies a branching ratio to the ground state that is consistent with the directly measured one shown in Fig.~\ref{fig2}d.  We also observe a consistent loss-detected fidelity of \FidelitygRBEchoLossdetected when including an echo (Methods), which suggests that the measured gate fidelity with loss detection is independent of the benchmarking method. This achieved gate fidelity is one of the highest reported to date in the neutral atom platform~\cite{tsai2025benchmarking,muniz2025high}; extensive modeling of the two-qubit gate errors (Extended Data Fig.~\ref{SMfig5} and Methods) suggests that residual unexplained errors are likely the result of imperfect optimization of the time-optimal gate.

We also characterize in more detail the loss during the entangling gates (Fig.~\ref{fig3}c). The loss per CZ gate extracted from the slope shows that the probability of correlated loss $p_\mathrm{corr}$ is orders of magnitude higher than the square of the probability of single-atom loss $p_\mathrm{single}^2$, which represents the expected probability of two-atom loss for uncorrelated single-particle loss. A simulation that assumes all the Rydberg decay leaving the qubit subspace goes to non-interacting states --- which, though simpler, is not entirely expected due to blackbody-induced decay to opposite parity states --- suggests that most of the correlated loss arises from instances where one atom decays, interrupts the blockade, and thereby causes the other atom to end up in the Rydberg manifold~\cite{wu2022erasure}. The remaining Rydberg state population is then converted to atom loss by the ionization beam.  With this mechanism, $p_\mathrm{corr}$ becomes a similar magnitude to $p_\mathrm{single}$. We note that quantitatively characterizing these errors could inform improved decoding in quantum error correction~\cite{baranes2025leveraging} as well as influence achievable code thresholds.

\begin{figure*}[t]
    \includegraphics[width=\textwidth]{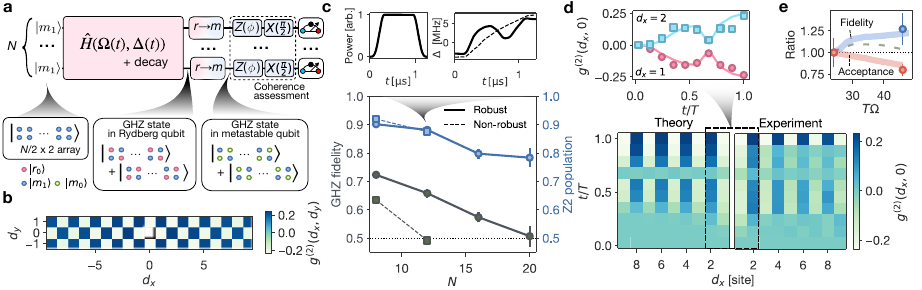}
    \caption{\label{fig4}%
    \textbf{Disorder-robust generation of $Z_2$-ordered GHZ state and mapping to the nuclear spin qubit.}
    \textbf{a,}~ Sequence of mapping and fidelity assessment. After the generation in the Rydberg state manifold, the GHZ state is mapped onto the non-interacting metastable qubit. The coherence is bounded by the phase scan of the global $\pi/2$-pulse as shown inside the dotted frame. \textbf{b,}~ Observed $Z_2$ correlation after the qubit mapping.  \textbf{c,}~ Scaling behavior of the GHZ-state fidelity in a two-dimensional ladder array after mapping to the $m$-qubit. Significant improvement is observed with the implementation of a pulse derived from a GRAPE optimization using a cost function considering position fluctuations. For $N=20$, we observe \FidelityGHZTwentyatomsZTwo $Z_2$ population and measure \FidelityGHZTwentyatoms GHZ-state fidelity. On top, pulse shapes for both robust (solid line) and non-robust (dashed line) pulses for $N=12$ are presented. \textbf{d,}~ Evolution of the correlation function. (Bottom) Evolution of the $d_x$ direction correlation function, comparing theory and experiment with loss detection. (Top) Comparison of the evolution between theory (solid line) and experiment (dots) at distances $d_x=1,2$. \textbf{e,}~ Loss-detected GHZ fidelity and acceptance rate as a function of the total pulse time. The blue dots show the relative change of the GHZ-state fidelity, while the red dots show the relative change of the acceptance rate. The solid lines represent the theory accounting for the loss detection. The dashed line represents the theoretical fidelity with Rydberg decay and without the loss detection, showing a peak due to the trade-off between adiabaticity and decay. 
    }
\end{figure*}

\subsection*{Generation and mapping of two-dimensional $Z_2$-ordered GHZ state}

The presented qubit mapping and loss detection translate to larger entangled states and many-body dynamics. We use this to prepare $Z_2$-ordered GHZ states in the $m$-qubit manifold. The entanglement is generated with an adiabatic sweep of the many-body Hamiltonian while the atoms are in the $r$-qubit manifold. We map the entangled state to the non-interacting $m$-qubit with a similar method described in Fig.~\ref{fig2}c.
The presence of the long-range interaction beyond the nearest neighbor results in a non-zero energy shift for $Z_2$-ordered states. We account for this shift by adding a detuning to the mapping pulse (Extended Data Fig.~\ref{SMfig7} and Methods). 
In Fig.~\ref{fig4}b, the observed density-density correlation function for the GHZ state of the mapped nuclear spin qubit is presented. The second order correlation function, $g^{(2)}(d_x,d_y)$, is defined as
$$
g^{(2)}(d_x,d_y) =\frac{1}{N_{d_x,d_y}} \sum_{i,j:\bm{r_i}=\bm{r_j}+(d_x,d_y)} \langle n_in_j\rangle-\langle n_i\rangle\langle n_j\rangle
$$ 
where $\hat{n}_i = \ket{r}_i\bra{r}_i$ and $N_{d_x,d_y}$ is the number of the atomic pairs whose position has the relation $\bm{r_i}=\bm{r_j}+(d_x,d_y)$. 
To evaluate the GHZ-state preparation fidelity, both population and coherence should be assessed~\cite{omran2019generation}. Our coherence measurement scheme uses global phase rotations as shown in Fig.~\ref{fig4}a. We assess the coherence by observing the evolution of the parity value while sweeping the global phase of the $\pi/2$-pulse (see Extended Data Fig.~\ref{SMfig7} and Methods).

We use optimal control, based on the Gradient Ascent Pulse Engineering (GRAPE) algorithm, to design a pulse for GHZ-state generation. The system Hamiltonian is written as,
$$
H/\hbar = \sum_i \left(\frac{\Omega(t)}{2}\hat{\sigma}_i^X-\Delta(t)\hat{n}_i\right)+\sum_{i<j}\frac{C_6}{|\bm{r}_i-\bm{r}_j|^6}\hat{n}_i\hat{n}_j
$$
where $\hat{\sigma}_i^X = \ket{r}_i\bra{m}_i+\ket{m}_i\bra{r}_i$ and $\bm{r}_i$ is the position of the atoms. The externally controllable global parameters $\Omega(t)$ and $\Delta(t)$ are Rabi frequency and laser detuning relative to the free-space atomic resonance, respectively. We adjust the lattice geometry to homogenize the interactions along each direction of the ladder, which compensates for the measured anisotropic atomic interactions (see Extended Data Fig.~\ref{SMfig6} and Methods). 
Unlike previous demonstrations using similar adiabatic schemes~\cite{omran2019generation}, we employ a two-dimensional ladder array. This geometry naturally realizes degenerate ground states of two $Z_2$-ordered states when the detuning, $\Delta$, is large, even without local light-shift on the outer edges~\cite{omran2019generation}. However, diagonal interaction introduces sensitivity to fluctuations in the atomic distance. This is because the strength of the diagonal interaction is on the order of \DiagonalEnergy,  comparable to the typical operating Rabi frequency \FreqRydbergRabiGHZ of the driving field. By a parameter sweep optimized without accounting for the shot-to-shot fluctuations of the interactions, we observe severe degradation of the GHZ-state coherence and consequently fidelity (Fig.~\ref{fig4}c). 

To mitigate this effect, the cost function for the GRAPE optimization is modified to explicitly incorporate interaction fluctuations (Methods). Interestingly, the optimized result shows a qualitatively different frequency sweep profile that is non-monotonic with time and crosses the phase transition point, $\Delta_c/\Omega \simeq 1.3$, three times (Extended Data Fig.~\ref{SMfig7_0} and Methods). Using this optimized pulse shape, we successfully generate GHZ states with up to 20 atoms.

To elucidate the physics behind this robust protocol, we show the evolution of the correlation function along the ladder direction, $g^{(2)}(d_x,0)$ in Fig.~\ref{fig4}d. 
A reduction of the correlation is observed in correspondence with the dip in the detuning profile, where the system transitions back into the trivial phase, followed by a revival toward the end of the pulse as it reenters the ordered phase. The dynamical phase accumulated during the intermediate trivial-phase regime plays a key role in coherently suppressing many-body phase errors induced by atomic position fluctuations, in direct analogy to the spin-echo effect in nuclear magnetic resonance, as further discussed in Ref.~\cite{zeng2025adiabatic}.

This interference mechanism is generic and can be applied to a broad class of interacting systems where static perturbations break a symmetry of the ideal Hamiltonian. In the present system, the symmetry-breaking perturbation arises from positional fluctuations of the atoms, which break the $Z_2$ symmetry characterizing the GHZ state. In Ref.~\cite{zeng2025adiabatic}, we further demonstrate its applicability to GHZ-state generation in quantum Ising models, as well as to $Z_3$ and $Z_4$ ordered states and quantum spin liquids in Rydberg arrays.

Detection of the Rydberg decay enables probing many-body dynamics even when there is a significant decay probability. To illustrate this feature, we compare the loss-detected fidelity of the GHZ states generated using two pulses, each with a different total pulse length. As the total pulse time, $T$, increases, we observe a trade-off: While the acceptance rate decreases as a result of increased Rydberg decay, the GHZ-state fidelity achieved improves. The enhancement results from a more adiabatic evolution of the longer pulse. Unlike the experiment on Fig.~\ref{fig2}e, the non-Hermitian contribution from Rydberg decay does not significantly affect the many-body dynamics, as its energy scale remains small compared to the energy scales of the Hamiltonian (see Extended Data Fig.~\ref{SMfig8} and Methods). In this regime, and at the expense of statistics, the loss-detected system evolves indistinguishably to a system with a Rydberg lifetime of \LifetimeRydbergSlow, a similar timescale to circular Rydberg states ~\cite{saffman2010quantum,wu2023millisecond}.

\subsection*{Outlook}

This architecture can be extended in several directions. The versatility of digital circuits incorporating non-local connectivity can be used for optimization of sensors such as programmable optical clocks~\cite{cao2024multi, finkelstein2024universal}, featuring error correction in state preparation, state read out (signal ``decoding"), or even during sensor operation~\cite{zhou2018achieving}, where all stages besides sensing occur in the robust nuclear qubit while sensing arises after mapping to the optical qubit~\cite{pedrozo2020entanglement,kaubruegger2021quantum,marciniak2022optimal,pezze2018quantum}. Such circuits could also be combined with stroboscopic shelving of optical coherence into the nuclear qubit for mid-circuit measurements on an atomic clock, to enable adaptive feedback.

At the same time, the mapping of Rydberg qubits to computational qubits allows for coherent atom rearrangement~\cite{bluvstein2022quantum} --- otherwise precluded by the Rydberg lifetime --- enabling direct access to entanglement witnesses of quantum criticality and topological order~\cite{pichler2016Measurement,ott2025probing}. Meanwhile, this mapping allows analog quantum simulations to be leveraged for generating resource states for quantum computation~\cite{martin2020digital,bauer2020quantum} and quantum metrology~\cite{eckner2023realizing,bornet2023scalable}. Conversely, mapping the computational qubit to the Rydberg qubit allows programmable entangled state preparation for subsequent many-body evolution. 

The high-fidelity decay detection enabled by our approach opens significant opportunities for both quantum simulation and quantum computing. 
When the Rydberg decay rate is larger than the characteristic energy scales of the Hamiltonian,  the capability to isolate no-decay events can reveal exotic phenomena such as entanglement propagation beyond the Lieb-Robinson bound~\cite{zhang2025observation,halati2025light}. On the other hand,  when the decay rate is smaller, we numerically find that the system behaves indistinguishably from the unitary evolution after post-selection. The achieved effective Rydberg lifetime could enable Rydberg based quantum simulations with unprecedented evolution time.
Moreover, high-fidelity control of the Rydberg transition via loss-detected error mitigation is particularly intriguing, as it eases the requirements for optical power and may contribute to scalable two-qubit gate implementation for neutral-atom quantum computers.

Finally, the use of GRAPE in many-body systems has enabled the discovery of a robust, experiment-tailored protocol that departs significantly from conventional monotonic adiabatic ramps. These results highlight the potential of optimal control techniques to uncover efficient strategies for steering strongly interacting systems and are expected to drive further advances in analog quantum simulation of many-body systems. For example, studies of quantum spin liquids~\cite{semeghini2021probing} might be advanced by combining robust sweep protocols~\cite{zeng2025adiabatic} with the error detection schemes reported here.

\ParaPlan{}

\bibliography{v3_arxiv2/theBib}

\clearpage

\section*{Methods}

\subsection*{Resource efficient atom preparation}

The total number of optical tweezers that can be generated is limited by available laser power. Further, without additional steps, only about $\sim50\%$ of these tweezers are occupied due to the stochastic nature of the loading process. To improve on this inefficiency and prepare fully occupied, defect-free arrays, we combine enhanced loading with rearrangement.

\Ybferm is loaded from the \WavelengthGreen narrow-line magneto-optical trap (MOT) to the \WavelengthTweezerOne tweezer array generated by acousto-optic deflectors (AOD) in a crossed configuration. Using the enhanced loading method reported in previous work~\cite{jenkins2022ytterbium}, we load single atoms with a \EfficiencyLoadingTypical probability. For the non-destructive imaging~\cite{lis2023midcircuit}, atoms are transferred to the \WavelengthTweezerTwo tweezer array generated by a liquid crystal spatial light modulator (SLM). Unlike previous experiments that used the \WavelengthClockMagic clock magic wavelength~\cite{ludlow2015optical}, we employ \WavelengthTweezerTwo tweezer during detections and coherent manipulations. 

After imaging, the atoms are rearranged by moving \WavelengthTweezerOne tweezers, using row-by-row compression. The resulting loading and rearrangement efficiency is shown in Extended Data Fig.~\ref{SMfig1}d. By combining rearrangement and enhanced loading~\cite{tian2023parallel}, we achieve \EfficiencySortingTypical defect-free array sorting success of up to \NumbersSortedArrayFraction atoms to the total tweezer array, such as a \NumbersSortedArraySize atom ladder out of a \NumbersArrayMax -tweezer loading array. Remarkably, even for a \NumbersSortedArraySizeMax atom array sorted out of \NumbersArrayMax total tweezers, we get a \NumbersSortedArrayFractionMax success rate. The performance of this sorting technique is summarized in Extended Data Fig.~\ref{SMfig1}.

\subsection*{Preparation of qubits in the metastable manifold} 

Our state preparation protocol builds on methods developed in previous work~\cite{lis2023midcircuit}. After loading and sorting the atoms, we apply Raman sideband cooling to bring them to the motional ground state in the radial directions (i.e., directions transverse to the tweezer propagation direction). Using sideband spectroscopy, we verify cooling to an average occupation number $\bar{n} < \NumbersNbarRaman$.

In the axial direction, the atoms are cooled only via gray molasses, as both the clock and Rydberg laser beams are applied transversely and do not couple strongly to axial motion. After Raman sideband cooling and optical pumping, we rotate the quantization magnetic field perpendicular to the initial magnetic field.

To prepare the desired spin state, we perform a spin flip from $m_F = +1/2$ to $m_F = -1/2$ in the ground-state manifold via a Raman rotation, followed by a $\sigma^+$-polarized clock pulse that excites the atom to the metastable clock state~\cite{jenkins2022ytterbium}. Importantly, we implement a state-preparation erasure protocol by applying resonant light onto the \ground$ \rightarrow $\onePone transition: Any error in the spin flip or clock excitation results in resonant scattering (preparation erasure) and atom loss (delayed erasure), allowing us to herald successful state preparation.

To ensure that the motional ground state is preserved during the clock excitation, we employ a motional-state-preserving pulse (MPP), as demonstrated in Ref.~\cite{lis2023midcircuit}. Using release-and-recapture thermometry on both ground and metastable states, we confirm that this excitation process introduces negligible heating (see Extended Data Fig.~\ref{SMfig1}g).

\subsection*{Motional state preserving pulse (MPP) in non-magic tweezer}

While \WavelengthClockMagic is the exact magic wavelength for the clock transition, generating high optical power at this wavelength is challenging and typically requires Ti:sapphire laser systems. For scalable architectures, operation at wavelengths accessible with fiber-laser technology is therefore preferable.

In our current tweezer system, operating at \WavelengthTweezerTwo generated by a frequency-doubled Er-doped fiber amplifier, we experimentally confirm performance comparable to that at the exact magic wavelength. This includes high-contrast $o$-qubit Rabi oscillations with over \NumbersRabiClockFlip coherent cycles~\cite{lis2023midcircuit}.

To evaluate the feasibility of MPP pulses across a broader range of wavelengths, we theoretically analyze their fidelity under varying levels of differential light shift inhomogeneity. We find that, provided the tweezer-induced light shifts remain sufficiently uniform across the atomic array, MPP pulses retain high fidelity even away from the exact magic wavelength.

Our simulation assumes a one-dimensional harmonic oscillator Hamiltonian
$$
H_\mathrm{total} = H_\mathrm{diag} +H_\mathrm{light}.
$$
After the rotating wave approximation, the diagonal part, $H_\mathrm{diag}$, consists of the energy levels of the harmonic oscillator states, as well as the laser detuning, $\Delta_m$, coming from the spatially varying differential light shifts due to the inhomogeneity of the traps:
$$
H_\mathrm{diag} = \sum_i\omega_g i \ket{g,i}\bra{g,i}+\sum_j(\omega_m j+\Delta_m) \ket{m,j}\bra{m,j} ,
$$
where $i,j$ are the labels for the motional levels. Here, it is assumed that the ground state, $g$, and the metastable clock state, $m$, have different trap frequencies $\omega_g$ and $\omega_m$. 

The optical excitation term is given by $H_\mathrm{light}$, where
$$
H_\mathrm{light} = \sum_{i,j}\Omega_m\bra{i}_{\omega_m}e^{i k x}\ket{j}_{\omega_g}\ket{m,i}\bra{g,j} + \mathrm{h.c.}
$$
The motional state matrix element for the non-magic condition is explicitly written as
$$
\bra{i}_{\omega_m}e^{i k x}\ket{j}_{\omega_g} = \int_{-\infty}^{\infty}dx\psi_i(x,\omega_m)^*e^{i k x}\psi_j(x,\omega_g) ,
$$
where $\psi_i(x,\omega)$ are the eigenstates of the one-dimensional harmonic oscillator with trap frequency $\omega$.

Using the polarizability data of \Ybferm~\cite{hohn2023state}, we numerically simulate this Hamiltonian, including up to the 15th motional level as shown in the Extended Data Fig.~\ref{SMfig1}g. The largest error contribution stems from the detuning caused by the tweezer inhomogeneity, and in the absence of inhomogeneity, the performance of the MPP is not significantly degraded by non-magic traps.

\subsection*{Fast destructive imaging and three outcome measurements}

For spin measurements as well as erasure detection, we employ destructive imaging using alternating counter-propagating beams resonant to the \ground $\longleftrightarrow$ \onePone transition, closely following the method demonstrated in~\cite{su2025fast}. We alternate the two beams with a frequency of \FreqBlueImagingAltering. For erasure detection, imaging is performed in \NumbersShallowTweezerDepth-deep tweezers used for clock and qubit operations, where the depth is quoted for the trapping potential of the ground state.

By selecting a sufficiently large region of interest (ROI) for each atom, we achieve clear separation between camera background counts and the single-atom fluorescence peak in photon counting histograms (Extended Data Fig.~\ref{SMfig2}c). A fit to two Gaussians indicates an erasure infidelity of \InfidelityBlueImagingErasure, after renormalizing the height to equalize the areas of the two peaks.

For spin detection in dense tweezer arrays, imaging in shallow tweezers leads to fast atomic diffusion, causing fluorescence to leak into the ROIs of neighboring tweezers and degrade spin state discrimination. To pin the atoms in place, we apply \NumbersDeepTweezerDepth-deep tweezers (Extended Data Fig.~\ref{SMfig2}d). Because the excited state \onePone is also trapped at \WavelengthTweezerTwo (Extended Data Fig.~\ref{SMfig1}c), this creates a deep confining potential even while the atoms scatter \WavelengthBlue photons at high rates~\cite{miranda2015site}. Using this approach, we achieve an infidelity of \InfidelityBlueImagingSpin, while still using individual ROI sizes smaller than half the tweezer spacing.

We can optimize the imaging time to minimize infidelity in dense arrays. We interpret this as a compromise between collecting sufficient fluorescence and minimizing cross-talk from neighboring atoms. As shown in Extended Data Fig.~\ref{SMfig2}e, we find an optimal imaging time near \TimeBlueImage. We note here that due to daily fluctuations in beam balancing and intensity, the ideal imaging time fluctuates and some datasets were measured with a time of \TimeBlueImageTwo.

Although the fast imaging itself takes only \TimeBlueImageOneTwo, the current total time for the three-outcome measurement is limited by the camera exposure, which is currently set by the duration of the initial non-destructive images. This could be improved in future iterations via software-controlled variable exposure. In that case, the limiting factor would become the tweezer ramp-up time prior to imaging, which is \TimeTweezerRampTimeImaging. While the destructive imaging is largely insensitive to atomic temperature, loss during the ramp imposes a constraint on overall imaging speed. Nevertheless, we project a total readout time below \TimeProjectedThreeOutcomeMeasurement for the full three-outcome sequence.

\subsection*{Loss and Error Budget of the Three-Outcome Measurement}
Extended Data Table~\ref{table1} presents the detection probabilities for each outcome of the three-outcome spin measurement. Each row corresponds to a different prepared qubit state, with the columns indicating: correct detections (diagonal), spin mislabeling errors (off-diagonal), and atom loss (rightmost column). While preparation erasure detection is applied, no correction is made for spin-flip errors introduced during the preparation of each input state.

The loss and infidelity observed in Table~\ref{table1} arise from two main sources: (1) errors during state preparation, and (2) errors introduced during the readout sequence. Table~\ref{table2} details the preparation error budget, listing for each step: (i) the atom loss probability, (ii) whether the error is detectable via erasure, and (iii) any associated spin-flip error. Most preparation-related loss arises from imaging and the tweezer ramp-down used for $o$-qubit operations. Spin-flip errors primarily result from polarization imperfections during the transfer into the metastable manifold, and from the $\pi$-pulse in the metastable manifold used to prepare $\ket{m_0}$. The total state preparation loss is 2.1(2)\% for $\ket{m_1}$ and 2.7(2)\% for $\ket{m_0}$, which reduce to 0.8(1)\% and 1.2(2)\%, respectively, after erasure correction.

Following state preparation, the three-outcome measurement sequence introduces additional sources of atom loss and spin mislabeling—defined as the incorrect assignment of a spin state (see Table~\ref{table3}). De-excitation from the clock state is performed using fast $\pi$-pulses in free space. The dominant contribution  arises from detuning errors caused by tweezer-induced light shifts (when operating in free space) and clock-light-induced shifts (from changing drive strengths throughout the sequence). These can, in principle, be mitigated by dynamically adjusting the laser frequency.

Spin mislabeling is primarily caused by spurious $\pi$-polarized components in the clock beam, which can drive unwanted transitions from $\ket{m_0}$ to $\ket{g_0}$. These polarization impurities may stem from imperfections in the optical beam or from misalignment of the magnetic field direction.

Destructive imaging is performed in deep tweezers, where the associated ramp-up of the tweezer potential induces additional Raman scattering. Other sources of error include imperfect spin rotations in the metastable manifold during state mapping and a finite probability of failing to identify an atom during imaging. Table~\ref{table3} distinguishes between loss and spin-mislabeling errors, and also indicates whether each error affects the first or second spin readout. Due to the sequential nature of the detection, some errors—particularly those affecting atoms in \mzero—are encountered twice, amplifying their impact.

While the spin mislabeling errors inferred from Table~\ref{table3} are in good agreement with the measured spin infidelities in Table~\ref{table1}, we observe a clear discrepancy in the total atom loss probability. This suggests the presence of an uncharacterized loss mechanism. Based on the larger discrepancy observed for the second spin readout \mzero, we hypothesize that the loss originates from the final ramp-up of the tweezer potential immediately preceding destructive imaging—a step encountered twice by \mzero atoms and not fully captured in individual calibration measurements we operated. Further investigation of this loss channel is left for future work.

\subsection*{Single-qubit operation in the metastable nuclear qubit}
Our single-qubit gate is operated with two different detunings depending on the applications. For the qubit mapping experiment, we choose a laser frequency positioned between the hyperfine lines of the excited \threeDone{} state to enable fast qubit rotations relative to the Rydberg dephasing rate. Under this condition, a Rabi frequency of \FreqMetaRabiFast{} is achieved due to constructive interference between the two Raman pathways.

For the two-qubit gate experiment, we use a slower Rabi frequency with an intermediate state detuning of approximately \DetuningMetaSlow{} from the $F=3/2$ transition to reduce both scattering from the intermediate state and errors associated with the finite turn-on time of the AOM across multiple pulses.

Single-qubit Clifford operations include a $Z$ gate implemented via phase tracking, taking advantage of the qubit energy splitting \FreqQubitSplit{}. Since this implementation is free from scattering error, the total scattering-induced error per Clifford gate is reduced relative to previous work~\cite{lis2023midcircuit}.

To assess the fidelity of our single-qubit operations, we perform randomized benchmarking (RB) using single-qubit Clifford sequences, as shown in Extended Data Fig.~\ref{SMfig3}c. For benchmarking the gates, we compare both erasure detection and loss detection schemes. As shown in Ref.~\cite{lis2023midcircuit}, scattering to the \Pone{} state (which decays to the \ground{} state) dominates the error mechanism in this method. Consequently, we observe improvements of similar magnitude using either erasure or loss detection.

The RB data are fit using the model $ap^l + b$, where $a$ and $p$ are fit parameters and $l$ is the number of gates. In the plot shown in Extended Data Fig.~\ref{SMfig3}, we assume $b = 0$ for the case without loss detection, and $b = 1/2$ for the case with loss detection. The gate infidelity is then extracted as $(1 - b)(1 - p)$.

Since the $Z$-gate is free from the scattering, we attribute most of the error to the $X$-gate where in our sequence we have \AverageXrot of $\pi/2$-pulses on average per single-qubit Clifford gate.

\subsection*{Measurement error correction for the Bell state}
Based on the characterization of the preparation and measurement errors, we can estimate the true Bell state fidelity~\cite{omran2019generation}. We define the measurement matrix
$$
M = \begin{pmatrix} 1-\epsilon_{00} &\epsilon_{10}\\ \epsilon_{01} &1-\epsilon_{11},
\end{pmatrix}
$$
for the measurement of a single spin, such that the relation between the measured count of each spin, $(N_0',N_1')$, and the actual count free from the measurement error, $(N_0,N_1)$, is 
$$
\begin{pmatrix} N_0'\\ N_1' \end{pmatrix} = M\begin{pmatrix} N_0\\ N_1 \end{pmatrix},
$$
 Here, we are not assuming conservation of counts such as $\epsilon_{00} =\epsilon_{01}$ since our measurement also induces loss. 

To correct for measurement error, we can infer the actual counts $(N_0,N_1)$ from the measured count $(N_0',N_1')$ by solving
$$
\min_{N_0,N_1}\norm{\begin{pmatrix} N_0'\\ N_1' \end{pmatrix} -M\begin{pmatrix} N_0\\ N_1 \end{pmatrix}}
$$

In the two-qubit experiments, we use $M\otimes M$ as a measurement matrix. Using this method, we analyze the metastable state Bell state data in Fig.~\ref{fig1}.
The measurement error corrected metastable Bell state fidelity is \FidelityBellMetaLossdetectedMcorr. The discrepancy to the two-qubit gate error is likely due to the loss-detected infidelity of single-qubit gate as well as \mone preparation infidelity.

We note that the dominant measurement error from the loss detectable measurement is not caused by the imbalanced detection probability, $\epsilon_{00},~\epsilon_{11}$, of each spin but rather by the spin mislabeling errors in the detection, $\epsilon_{01},~\epsilon_{10}$. Indeed, if we assume $\epsilon_{10}=0,\epsilon_{01}=0$ in the analysis above, the corrected Bell fidelity is \FidelityBellMetaLossdetectedMnoflip, very similar to the value before the correction, \FidelityBellMetaLossdetected.

\subsection*{UV system and $T_2^*$ coherence of the Rydberg qubit}

For coherent manipulation of the \WavelengthUV{} single-photon transition to the Rydberg state, we use a laser system based on sum-frequency generation (SFG) and second-harmonic generation (SHG)~\cite{schine2022long,wilson2011750}. An overview of the system is shown in Extended Data Fig.~\ref{SMfig4}a.

The \WavelengthUVFundamentalOne{} light is generated by a home-built interference filter diode laser (IFDL), which is locked to an ultra-low expansion (ULE) cavity with finesse around \NumbersFinesseULERydberg{}. To suppress servo bumps induced by current feedback, we amplify the cavity transmission via an injection lock~\cite{levine2018high},  whose output is further amplified by a Yb-doped fiber amplifier. The \WavelengthUVFundamentalTwo{} telecom wavelength is generated by a commercial fiber seed laser and subsequently amplified by an Er-doped fiber amplifier. Both fiber amplifiers can deliver up to \PowerUVSystemFiberAmplifiersMax{} of optical power.

The two amplified beams are combined in a periodically poled lithium niobate (PPLN) crystal for SFG. A portion of the generated \WavelengthUVSFG{} beam is sent to a ULE cavity to stabilize the frequency of the \WavelengthUVFundamentalTwo{} seed laser. After a power stabilization using an acousto-optic modulator (AOM), \PowerUVSystemSFG{} of \WavelengthUVSFG{} light is coupled into an SHG cavity. The SHG cavity contains a cesium lithium borate (CLBO) crystal, heated in an oven. This produces \PowerUVSystem{} of 302\nm{} UV light. We choose CLBO over beta barium borate (BBO) due to its smaller walk-off angle, which results in improved output beam quality~\cite{lorenz2021rydberg}. To mitigate crystal degradation, the SHG cavity is unlocked after each experimental run and re-locked during the initial atom preparation stage.

To deliver light from the SHG cavity to the atoms with high stability, we implement several measures. A hydrogen-loaded UV fiber is used to suppress pointing fluctuations originating from the SHG cavity and upstream AOMs~\cite{marciniak2019design}. We also implement an intensity servo that stabilizes the UV power via an AOM that deflects excess light. Additionally, a second intensity feedback loop is applied at the final AOM using a sample-and-hold method to stabilize the pulse amplitude immediately before use. 

Eight in-vacuum electrodes are used to control the electric field at the location of the tweezer array. To null the residual field, we minimize the Stark shift of the Rydberg state. Coherence between the laser and the Rydberg transition is characterized via Ramsey-type experiments (Extended Data Fig.~\ref{SMfig4}b), yielding an average $T_2^*$ coherence time of \TimeRamsey{} across the array. The observed electric field gradient is \NumbersEFieldGradient{}, which is small enough to have minimal impact on gate and simulation experiments, where typical energy scales are on the order of MHz.

Extended Data Fig.~\ref{SMfig4}e shows Rydberg Rabi oscillations of a single atom using this system. With loss detection, we observe \NumbersRabiSingleFlip{} coherent Rabi oscillations before $1/e$ decay, representing the highest number of coherent spin flips observed to date in Rydberg qubits~\cite{madjarov2020high}. When two atoms are placed close enough for the Rydberg blockade effect, the system oscillates between $\ket{mm}$ and the entangled state $(\ket{mr} + \ket{rm})/\sqrt{2}$. While the blockade enhances the Rabi frequency by a factor of $\sqrt{2}$, we observe loss-detected coherent cycles of \NumbersRabiBlockadedFlip{} ---comparable to the single-atom case--- indicating that the cycle number is mainly limited by shot-to-shot intensity fluctuations~\cite{madjarov2020high}.

For the $r$-qubit measurement without loss detection shown in Fig.\ref{fig1}, we employ auto-ionization using an inner-shell atomic transition at \WavelengthIonization{}~\cite{burgers2022controlling}. A beam of power \NumberIonPower{} is focused to a \NumberIonWasit{} Gaussian waist, producing an ionization decay time constant of \NumberIonDecay{}, fast enough to suppress Rydberg decay prior to ionization~\cite{madjarov2020high}.

\subsection*{Detection scheme of the Rydberg qubit} While in most measurements of this work, we detect the Rydberg qubits using mapping to the $m$-qubit and subsequently conducting the loss detectable spin measurement, some of the data in Fig.~\ref{fig1}f relies on the direct detection of the $r$-qubit spin states. This was done by ionizing the Rydberg state via an auto-ionization beam~\cite{madjarov2020high,burgers2022controlling}. The ionized atoms are not trapped anymore, leave the trap and get counted as a loss while the atoms in the \mone state remain in the trap and eventually get imaged. As we show in the Extended Data Fig.~\ref{SMfig4_5}, our ionization rate is \NumberIonDecay which suppresses the Rydberg decay error during the measurement to less than \NumberIonDecayErrLim.

\subsection*{Rydberg state decay branching ratios} We estimate the branching ratios of the Rydberg state from various experiments shown in Extended Data Fig.~\ref{SMfig4_5}b,c. For estimating the decay to the ground state, \ground, as well as the \mone and \mzero metastable state, we fitted the initial population accumulation speed with a linear fit and compared to the speed of the loss of the total Rydberg state population. To mitigate the effect of atomic loss as well as to stay in the linear part of the initial slope of the decay, we only fit the first 30 \us~ of the data. The branching ratio to the  $^3\mathrm{P}_2~F=3/2$ state is measured in a separate experiment where we compared the decayed amount to the ground state with and without the $^3\mathrm{P}_2~F=3/2\rightarrow {}^3\mathrm{S}_1~F=1/2$ repumper. Out of the ratio of those measured rates as well as the \BranchThreeSoneToThreePzero decay probability to the \clock state, the decay to the $^3\mathrm{P}_2~F=3/2$ state is identified to be \RydbergBranchthreePtwoRatio of the decay to the \ground state. Using this information we extract the final branching ratio of \RydbergBranchPop for \ground, $^3\mathrm{P}_2~F=3/2$, \mone, \mzero state, remaining with an undetected population of \RydbergBranchOtherPop. This includes the decayed portion to the other hyperfine state of $^3\mathrm{P}_2$, $F=5/2$.

\subsection*{Two-qubit gate calibration and error analysis}

For empirical optimization of the two-qubit gate, we use one of the 20 CZ gate sequences with spin-echo pulses inserted between each pair of CZ gates as a target sequence, following Ref.~\cite{evered2023high}. We benchmark the resulting gate using echo-type global randomized benchmarking (gRB), extracting a loss-detected gate fidelity of \FidelitygRBEchoLossdetected{}, which is consistent with the non-echoed result presented in Fig.~\ref{fig3}.

For the non-echoed sequence, we compensate the single-qubit phase accumulated during gate operation using an additional $Z$ rotation. This compensation is optimized by maximizing the return probability of one instance of the 20-CZ non-echoed gRB sequence. For both echo and non-echo gRB, we use \NumbersgRBInst{} random instances per two-qubit gate depth $l$.

To analyze the benchmarking data, we fit the success probability using the function $ap^l + b$, where $a$ and $p$ are fit parameters and $l$ is the number of CZ gates. The asymptotic parameter $b$ is chosen based on the detection scheme. For gRB with erasure preparation and erasure decay detection, we assume $b = 0$ since leakage from the qubit subspace dominates and results in vanishing success probability in the infinite-depth limit. For loss detection, we set $b = 1/4$, as only one of the four computational basis states $\{00, 01, 10, 11\}$ is observed per run~\cite{muniz2025high,radnaev2025universal}. The error per gate is then extracted as $(1 - b)(1 - p)$.

To understand the sources of gate error, we perform a master equation simulation using a $16 \times 16$ density matrices with 4-levels per atom, including the states $\ket{m_0}$, $\ket{m_1}$, $\ket{r_0}$, and an auxiliary decayed state $\ket{\mathrm{decay}}$. This state captures decay from $\ket{r_0}$ to levels outside the qubit manifold. The decay is modeled with operators $\sqrt{\Gamma_\mathrm{eff}} \ket{\mathrm{decay}}\bra{r_0} \otimes I$ and $I \otimes \sqrt{\Gamma_\mathrm{eff}} \ket{\mathrm{decay}}\bra{r_0}$ for each atom, where $\Gamma_\mathrm{eff}$ is derived from the branching ratio measured in Fig.~\ref{fig2}. Decay to $\ket{m_0}$ and $\ket{m_1}$ is also included accordingly, adding up to 6 decay operators in total. This simulation scheme is also applied to the simulation of the correlated loss effect in Fig.~\ref{fig3}.

All other imperfections are modeled as coherent errors and averaged over 1000 noise realizations. We include the following noise sources~\cite{shaw2024learning}:

\textbf{AC intensity}: Laser intensity fluctuations as time-varying Rabi frequency extracted from relative intensity noise (RIN).

\textbf{DC intensity}: Laser shot-to-shot intensity fluctuations as shot-to-shot Rabi frequency fluctuation sampled from gaussian noise. The standard deviation is estimated from photo-detector measurements.

\textbf{Beam pointing}: Laser beam pointing fluctuations are modeled as shot-to-shot Rabi frequency fluctuations sampled from Gaussian noise. The magnitude of this behavior is derived from spatial jitter observed in camera images.

\textbf{Beam sampling}: Atom wavepacket's finite width to the perpendicular direction of the UV laser as shot-to-shot Rabi frequency fluctuations sampled from Gaussian noise. The standard deviation is estimated from the axial temperature and tweezer width.

\textbf{AC phase}: Laser frequency noise. Obtained from the PDH error signal of the \WavelengthUVSFG{} laser, corrected for the ULE cavity roll-off and SHG cavity transmission characteristics, as well as by the noise scaling of the SHG.

\textbf{Doppler}: Error due to finite velocity of the laser direction modeled as a shot-to-shot detuning fluctuation. The standard deviation is estimated from the radial atomic temperature.

\textbf{DC electric field}: Constant detuning errors due to measured electric field gradient. To deduce the gate-error result averaged over the array, detuning is sampled from a uniform distribution in the range observed in Fig.~\ref{SMfig4}

To assess the expected gate fidelity, we simulate its performance on 1000 instances of $m$-qubit Haar-random input states. For the loss-detected case, fidelities are computed after renormalizing the state within the $m$-qubit manifold. The results are shown in Extended Data Fig.~\ref{SMfig5}, where we find a total error of \InFidelityCZTheory{} without loss detection and \InFidelityCZTheoryLossdetected{} with loss detection. While we also examined the effect of tweezer position fluctuation, we find in the highly-blockaded regime, the related position error is several orders lower than the other quoted errors, therefore omitting it from the Figure.

The resulting theory values deviate by approximately \InFidelityCZTheoryExpDescrepancy{} from experiment for both before and after loss detection. We hypothesize that the remaining discrepancy is likely due to imperfect convergence of the empirical optimization. In the theoretical modeling of the optimization, we numerically observe slow convergence of the gate fidelity at the 0.1\% level. We leave further investigation of this effect to future work.

\subsection*{Population dynamics during non-Hermitian evolution}

We consider the non-Hermitian state evolution in presence of Rydberg decay $\Gamma$. Initially we prepare the state $\sqrt{1-|a|^2}|m\rangle+a|r\rangle$, which evolves to the state after non-Hermitian evolution:
\begin{equation}
\psi_\mathrm{non-H}=\frac{\sqrt{1-|a|^2}|m\rangle+ae^{-t\Gamma/2}|r\rangle}{\sqrt{1-|a|^2(1-e^{-t\Gamma})}}
\end{equation}
Converting this equation using the initial Rydberg state population $p_0=|a|^2$, the effective Rydberg state population in the non-Hermitian evolution is,
\begin{equation}
P_{r}(t)=\frac{p_{r0}e^{-t\Gamma}}{(1-p_{r0})+p_{r0}e^{-t\Gamma}}
\end{equation}

\subsection*{Calibration of the anisotropic Rydberg interaction}

To engineer the correct Hamiltonian in our quantum simulation experiments, precise calibration of the interaction is crucial. To perform this calibration, we directly measure the interaction between atoms using a two-photon transition to the $\ket{rr}$ state from the non-interacting $\ket{mm}$ state, with the {$(\ket{mr}+\ket{rm})/\sqrt{2}$ state as an intermediate state of the two-photon process. The detuning of such a two-photon transition resonance from the free space resonance corresponds to half of the interaction energy. 

With this method, as predicted in~\cite{peper2025spectroscopy}, anisotropy of the Rydberg interaction is observed, depending on the inter-atomic orientation relative to the quantizing magnetic field (Extended Data Fig.~\ref{SMfig6}). To achieve an effectively uniform interaction, we adjust the lattice aspect ratio by \NumbersRydbergLatticeDeformation.

\subsection*{GHZ-state mapping and staggered magnetism}
To map the GHZ state from the interacting $r$-qubit manifold to the non-interacting $m$-qubit manifold, we use two consecutive global $\pi$ pulses on the $m$-qubit and on the $r$-qubit, as described in Fig.~\ref{fig2}~\cite{bluvstein2022quantum}. For the second $\pi$ pulse, we account for the resonance shift due to interactions between Rydberg states using a detuning of \NumberProjectionDetuning (Extended Data Fig.~\ref{SMfig7}).

In Fig.~\ref{fig1}, we use the quantity of staggered magnetism to assess the $Z_2$-ordered population~\cite{omran2019generation}. The staggered magnetism for the ladder geometry is defined as $M_\mathrm{stagger} =\sum_{i=1,..,20} (-1)^i\langle \sigma^Z_i\rangle$, where the index $i$ is defined in a circular manner around the ladder geometry and the $\sigma^Z=\ket{0}\bra{0}-\ket{1}\bra{1}$ is defined in the $r$-qubit. From this definition, $|M_\mathrm{stagger}|=N$ only occurs when the array is showing the complete $Z_2$-order, and we find a $Z_2$-population summing up $M_\mathrm{stagger} =\pm N$ populations.

\subsection*{Parity measurement of GHZ-state coherence}

The GHZ fidelity of a state $\hat{\rho}$ reads
\begin{equation}
F=\langle\mathrm{GHZ}|\hat{\rho}|\mathrm{GHZ}\rangle
=\tfrac12\bigl(\rho_{AA}+\rho_{\bar A\bar A}\bigr)
+\mathrm{Re}\bigl(\rho_{A\bar A}\bigr),
\nonumber
\end{equation}
where $|A\rangle$ and $|\bar{A}\rangle$ are the two checkerboard configurations, and $\rho_{\alpha\beta}=\langle\alpha|\hat{\rho}|\beta\rangle$. 
To infer the coherence term $\mathrm{Re}\bigl(\rho_{A\bar A}\bigr)$ following the mapping to the $m$-qubit manifold, we measure the parity operator $\hat{\Pi} = \prod_i \hat{\sigma}^Z_i$ after applying a global shift $\hat{U}(\phi) = \exp(- i \phi \sum_i \hat{n}_i )$ followed by a $\pi/2$ rotation $\hat{U}_x = \exp( -i \frac{\pi}{4} \sum_i \hat{\sigma}^X_i  ) $.
As we show below, with this method, we can extract $\mathrm{Re}\bigl(\rho_{A\bar A}\bigr)$ from the offset of the parity oscillation rather than the amplitude. This approach differs from the standard parity-based coherence measurements requiring local control~\cite{omran2019generation}, and generalizes the global-control method of Ref.~\cite{wilk2010entanglement} to many-body systems.

The measured parity is given by
\begin{align}
\nonumber
\Pi(\phi) & = \text{Tr}(\hat{U}_{x}\hat{U}(\phi) \, \hat{\rho}\, \hat{U}(\phi)^{\dagger}\hat{U}_{x}^{\dagger} \, \hat{\Pi}) = \\
\nonumber
& =  \sum_{n} (-1)^{N_{n} + \frac{N}{2}}\rho_{n\bar{n}}e^{-i \phi (N_{n}-N_{\bar{n}})},
\end{align}
where $N_n$ is the number of Rydberg excitations in the configuration $n$, and $\bar{n}$ is the configuration where all spins in $n$ are flipped.
By taking the average over $\phi$ we get rid of the oscillatory terms with $N_n\neq N_{\bar n}$ and isolate the off-diagonal $\hat{\rho}$ components with $N_n=N_{\bar n}$:
\begin{equation}
\overline{ \Pi(\phi) } = 2\,\mathrm{Re}(\rho_{A\bar A}) + \sum_{m\in S_A} \rho_{m\bar m},
\nonumber
\end{equation}
where $S_A$ is the set of all configurations obtained by permuting Rydberg excitations in $A$.
From this equation, we obtain a bound for the GHZ coherence:
\begin{align}
2 \mathrm{Re}(\rho_{A\bar{A}})
&= \overline{ \Pi(\phi) } - \sum_{m \in S_A} \mathrm{Re}(\rho_{m\bar{m}})
\nonumber \\
&\geq \overline{ \Pi(\phi) } - \sum_{m \in S_A} |\rho_{m\bar{m}}|
\nonumber \\
&\geq \overline{ \Pi(\phi) } - \sum_{m \in S_A} \sqrt{ P_{m} \, P_{\bar{m}} } \,,
\label{eq:coherence_lowerbound}
\end{align}
where we used $|\mathrm{Re}(x)|\le|x|$ and the Cauchy–Schwarz inequality $|\rho_{m\bar m}|^2\le P_{m}\, P_{\bar m}$, with $P_m$ the population of configuration $m$.  

When evaluating the parity average $\overline{\Pi(\phi)}$ using a finite number of sampling points for $\phi$, one must consider that $\Pi(\phi)$ contains oscillatory contributions at various frequencies. These arise from terms with configurations $n$ and $\bar{n}$ such that $N_n - N_{\bar{n}} = \delta N \ne 0$, leading to oscillations with period $T = 2\pi  \delta N$. The corresponding amplitude is 
\begin{equation}
A_{\delta N} = \sum_{ |N_n - N_{\bar{n}}| = \Delta N} \! \! \! |\rho_{n\bar{n}}|
\leq \sum_{|N_n - N_{\bar{n}}| = \delta N} \sqrt{P_n P_{\bar{n}}} ,
\nonumber
\end{equation}
where we used the Cauchy–Schwarz inequality in the final step. From this expression, we see that oscillations with $\delta N \ne 0$ are strongly suppressed. This is because if $N_n < N/2$, then necessarily $N_{\bar{n}} > N/2$, or vice versa. As a result, at least one of the configurations $n$ or $\bar{n}$ must locally violate the Rydberg blockade, leading to a significantly reduced probability $P_n \ll 1$. The suppression becomes more severe with increasing $\delta N$, since larger imbalances require increasingly incompatible excitation patterns. For the GHZ state prepared using the optimal robust pulses in Extended Data Fig.~\ref{SMfig7_0}, we estimate that the amplitude of these oscillations is small ($< 10^{-2}$) even for systems of 20 atoms. 

Assuming that only the $\delta N = 2$ components contribute to the oscillations beyond the offset $\overline{\Pi(\phi)}$, any uniform sampling grid in $\phi$ over a $2\pi$ interval is sufficient to exactly extract the parity offset. In the experiment, we sample 11 equally-spaced values of $\phi$ between 0 and $2\pi$ (Extended Data Fig.~\ref{SMfig7}e). The observed dependence of $\Pi(\phi)$ on $\phi$ in the experimental data can be attributed to two main factors: shot-to-shot geometry variations due to the fluctuations in the atomic positions that are included in our numerical simulations and discussed below, and imperfections in the overall parity measurement procedure and in the GHZ preparation that are not captured by the simulations.

In Extended Data Fig.~\ref{SMfig7}g, we plot the exact GHZ coherence $2 \mathrm{Re} \left( \rho_{A \bar{A}} \right)$ for a 16-atom GHZ state, prepared using the robust pulse shown in Extended Data Fig.~\ref{SMfig7_0}, as a function of the GHZ coherence estimated from the bound in Eq.~\eqref{eq:coherence_lowerbound}. Different markers correspond to independent realizations of the fluctuating atomic positions $\mathbf{r}_i + \delta \mathbf{r}_i$, where $\delta \mathbf{r}_i$ are drawn from a gaussian distribution with zero mean and standard deviation $\delta r = 74\,\mathrm{nm}$, which is approximately 2\% of the lattice spacing. In Extended Data Fig.~\ref{SMfig7}h, we show the estimated GHZ coherence as a function of $\delta r$, for $\delta r$ up to $111\,\mathrm{nm}$. For each disordered instance, the lower bound in Eq.~\eqref{eq:coherence_lowerbound} is nearly saturated, indicating that the parity offset serves as a reliable estimator of GHZ coherence. However, positional disorder—though partially mitigated by the robust pulses—still induces significant shot-to-shot fluctuations in the exact GHZ coherence, accounting for the large experimental error bars.

\subsection*{Optimal control for GHZ-state preparation}

The GHZ-state preparation pulse is optimized using the Gradient Ascent Pulse Engineering (GRAPE) algorithm~\cite{Khaneja2005}. The control parameter $\Delta(t)$ is discretized on a time grid of $N=150$ points $t_j = \left( j + \frac{1}{2} \right) dt$, $j = 0,1,\dots,N-1$, with $dt = T/N$, and $T$ the total evolution time, such that $\Delta_j = \Delta(t_j)$ are the variational parameters of the optimization. The Rabi frequency $\Omega(t)$ is fixed to a cosine-tapered window function (cf.~Fig.~\ref{fig4}c). The cost function is defined as the preparation infidelity averaged over $N_s$ realizations of the fluctuating geometry:
\begin{equation}
    \mathcal{C} = 1 - \frac{1}{N_s} \sum_{k=1}^{N_s} \left| \langle {\rm GHZ}  | \prod_{j=0}^{N-1}  e^{-i dt \hat{H}^{(k)}(\Delta_j) } | m m \dots m \rangle \right|^2,
\end{equation}
where $\hat{H}^{(k)}(\Delta_j)$ is the many-body Rydberg Hamiltonian with atoms at positions $\mathbf{r}_i + \delta \mathbf{r}^{(k)}_i$, with $\mathbf{r}_i$ are the ladder coordinates. To ensure the smoothness of the optimal control function $\Delta(t)$, we add a penalty term to the cost function: $C \rightarrow C + \eta \int_{0}^{1} d s \, (d \Delta/ds)^2$, where $s = t/T$ and $\eta = 10^{-3}$.
In the numerical optimization, $N_s = 30$ and $\delta \mathbf{r}^{(k)}_i$ are drawn from a gaussian distribution with zero mean and standard deviation $\delta r = 60 \nm$. We verified that the resulting optimal pulses are independent of the number of samples and on $\delta r$.
The optimization is performed using a gradient-based algorithm, where the gradient can be efficiently computed from the approximation $\partial \hat{U}_j /\partial \Delta_j \simeq - i dt  \, \partial \hat{H}(\Delta_j)/\partial \Delta_j \, \hat{U}_j $. The optimization process begins with a short total evolution time $T$ and a linear profile for $\Delta(t)$ as the initial guess. Upon convergence, $T$ is incrementally increased as $T \rightarrow T + d T$, and the optimized profile at total time $T$ is used as the initial guess for the optimization at total time $T + dT$. 

\subsection*{Effect of the Rydberg decay detection}

Our decay-detection capabilities enable us to infer an effective Rydberg lifetime of 
$\LifetimeRydbergSlow$. 
However, it is not a priori guaranteed that the post-selected dynamics 
(conditioned on no detected decay events) is equivalent to the dynamics of a system 
with such an extended lifetime. To clarify this, we compare the ideal unitary 
(no-decay) evolution with the post-selected (no-jump) dynamics, which is governed by the 
effective non-hermitian Hamiltonian
$H = H_0 - \frac{i}{2}\sum_jn_j$, where the state is normalized after each time step. 
As shown in Extended Data Fig.~\ref{SMfig8}, the resulting no-jump dynamics 
reproduces the ideal unitary evolution with high accuracy even for decay rates 
up to an order of magnitude larger than the experimental condition, confirming that 
the observed dynamics faithfully represents coherent GHZ-state preparation. In the same figure, we also explore the quench dynamics under a time-independent Hamiltonian at zero Rydberg detuning, showing larger but still limited deviations between the no-jump and ideal unitary dynamics under the present experimental conditions.

We note, however, at much larger decay rates the no-jump dynamics 
deviates significantly from the purely unitary case.
In this regime, 
experiment reflecting the feature of non-Hermitian quantum dynamics can be realized, such as enhanced evolution of the correlation, 
provided that decay rates can be artificially engineered and sufficient 
post-selection statistics are maintained~\cite{zhang2025observation,halati2025light}.

\subsection*{Error bars and fitting}
Error bars on populations and parity measurements are 68\% Clopper-Pearson confidence intervals. Fits of the experimental data are done using weighted least squares and error bars on fitted parameters represent one standard deviation fit errors.

\subsection*{Acknowledgements}
\noindent
We acknowledge discussions with A. Cao, G. Liu, M. Peper, and J. D. Thompson; technical input from W. Eckner, C. Gross, P. Osterholz, F. Rönchen F. Vietmeyer, A. Wilson;  and, comments on the manuscript from N. Darkwah-Oppong, J. Thompson, D. Young and Y. Zhan.

A.B. is supported by the Swiss National Science Foundation under grant 222216. 
G. G. acknowledges support from the European Union’s Horizon Europe program under the Marie Sk\l{}odowska Curie Action TOPORYD (Grant No. 101106005).
These results are based upon work supported by the Office of Naval Research (N00014-23-1-2533), Air Force Office of Scientific
Research (FA9550-23-1-0097), Army Research Office / LPS (W911NF24S0004),  U.S. Department of Energy, Office of Science, National Quantum Information Science Research Centers, Quantum Systems Accelerator, NSF Convergence Accelerator, the NSF QLCI Award OMA - 2016244, Physics Frontier Center PHY-2317149, the National Institute of Standards and Technology, the ERC Starting grant QARA (Grant No. 101041435), the Horizon Europe programme HORIZON-CL4-2022-QUANTUM02-SGA via the project 101113690 (PASQuanS2.1) and by the Austrian Science Fund (FWF) (Grant No. DOI 10.55776/COE1). 

\noindent\emph{Note}: While completing this work, we became aware
of complementary work on erasure enhanced quantum error correction utilizing ${}^{171}$Yb metastable qubit ~\cite{zhang2025leveraging}.

\subsection*{Author Contributions}
\noindent

A.S., A.B., J.W.L., G.M.V., and A.M.K. contributed to the experimental setup, performed the measurements, and analyzed the data. Z.Z., G.G., and H.P. theoretically developed the disorder-robust pulses and the GHZ-state coherence measurement method with global control. A.M.K. and H.P. supervised the work. All authors contributed to the manuscript.

\subsection*{Competing Interests}
\noindent
The authors declare no competing interests.

\subsection*{Materials and Correspondence}
\noindent
Correspondence and requests for materials should be addressed to A.M.K.

\subsection*{Data Availability}
\noindent
The data that support the findings of this study are available from the corresponding
author upon reasonable request. Source data for figures 1--4 are provided with the paper.

\setcounter{figure}{0}
\renewcommand{\figurename}{Extended Data Fig.}
\renewcommand{\tablename}{Extended Data Table.}

\begin{figure*}
    \centering
    \includegraphics[width=\textwidth]{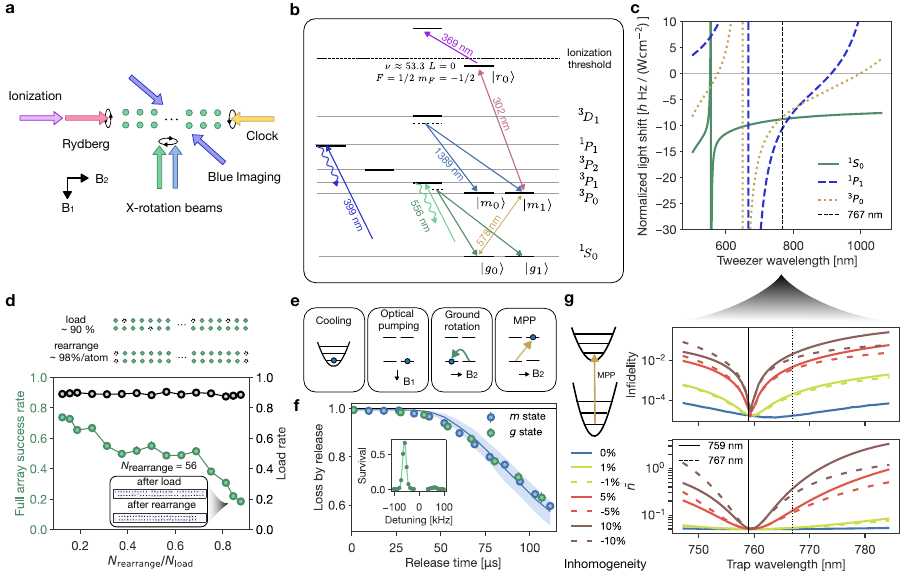}
    \caption{\label{SMfig1}%
\textbf{Beam geometry, atomic structure, tweezer polarizability, resource-efficient rearrangement, and state preparation.}
\textbf{a,}~Beam and magnetic field geometry relative to the atom array. $B_1$ is used during initial optical pumping and Raman cooling, followed by $B_2$ for the qubit manipulations. The X-rotation beams for both the ground and metastable states propagate perpendicular to $B_2$. The Rydberg and clock beams are aligned parallel to the magnetic field and use circular polarization to drive the $\sigma^+$ transition. The ionization beam is co-aligned with the UV beam.
\textbf{b,}~Atomic structure of \Ybferm showing the transitions relevant to the experiment. The \onePone{} transition is used for fast destructive imaging, while \Pone{} serves as both the non-destructive imaging transition and the intermediate state for Raman transitions. The metastable state is coupled to the Rydberg state via a single-photon transition. The ionization beam is resonant with an inner-shell transition.
\textbf{c,}~Polarizability of excited states compared to the ground state \ground. At the operational tweezer wavelength \WavelengthTweezerTwo{}, the \clock{} state is nearly magic, and the \onePone{} state used for destructive imaging remains trapped. For \onePone{}, only the scalar polarizability is considered due to scattering from both magnetic sublevels.
\textbf{d,}~Enhanced loading and rearrangement. We use single-atom loading at an efficiency of \EfficiencyLoadingTypical{} followed by rearrangement to prepare defect-free arrays, as illustrated in the top graphic. The data show the full-array success rate as a function of target array size.
\textbf{e,}State preparation sequence to the metastable state. After preparing the radial motional ground state via Raman sideband cooling, we flip the spin in the ground-state manifold using a $\pi$-pulse. Motional-state-preserving pulses (MPPs) are used to coherently excite the atoms to the \clock{} state without adding motional excitation\cite{lis2023midcircuit}.
\textbf{f,}~Measurement of atomic temperature. Inset: sideband spectroscopy indicating $\bar{n} = \NumbersNbarRaman{}$. Main panel: release-and-recapture comparison between ground and metastable states. The agreement shows that the clock excitation does not introduce additional motional heating. The solid line shows the Monte-Carlo simulation best agrees with the experiment. We extract the atomic temperature of \NumbersTempMonteCarlo.
\textbf{g,}~Simulation results for motional-state-preserving pulses at varying wavelengths. The upper panel shows the population transfer infidelity; the lower panel shows the motional excitation added by the pulse, indicating heating.}
\end{figure*}

\begin{figure*}
    \centering
    \includegraphics[width=\textwidth]{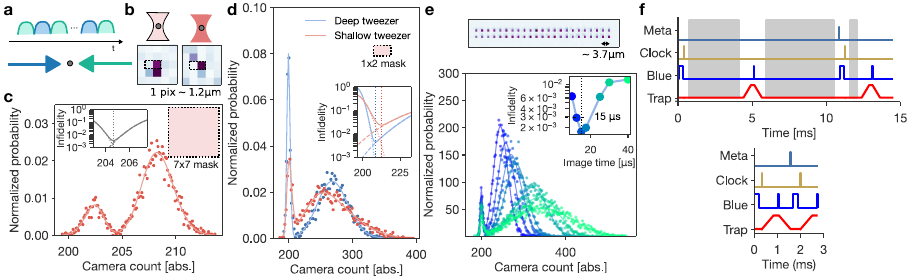}
    \caption{\label{SMfig2}%
\textbf{Blue imaging and experimental sequence for the three-outcome measurement.}
\textbf{a,}~Beam geometry and sequence of the destructive imaging. Counter-propagating beams resonant with the \ground{} to \onePone{} transition are alternately applied on the atoms at a frequency of \FreqBlueImagingAltering{}.
\textbf{b,}~Two methods of fast imaging. We use either \NumbersShallowTweezerDepth{}-deep tweezers (``shallow") or \NumbersDeepTweezerDepth{}-deep tweezers (``deep") in the experiment. The shallow tweezers are used for erasure detection, while the deep tweezers are used for spin detection. Bottom: typical single-shot images from both methods. The deep tweezers confine the atom position more tightly.
\textbf{c,}~Photon count histogram from shallow-tweezer imaging using a $7 \times 7$ pixel region of interest (ROI). A two-Gaussian fit infers a spin-readout infidelity below \InfidelityBlueImagingErasure{}.
\textbf{d,}~Comparison of photon count histograms from deep and shallow tweezers using a small ROI. The deep tweezers result in reduced overlap due to better confinement.
\textbf{e,}~Optimization of imaging time in a tightly spaced array matching the GHZ experiment geometry (Top). As imaging time increases, infidelity initially decreases due to improved fluorescence collection, then increases beyond \TimeBlueImage{} due to cross-talk from neighboring atoms. A similar trend is observed in the geometry for the gate experiment, which uses even smaller atom spacing (\LengthscaleGateDistance{}).
\textbf{f, }Three-outcome imaging sequence. ``Meta'' represents the $\pi$-pulse on the $m$-qubit subspace. The $\pi$-pulse on the $o$-qubit is indicated as ``Clock''. ``Blue'' indicates application of the resonant beam to the \ground{}$\longleftrightarrow$\onePone{} transition, used for both, imaging and blow-away of ground-state atoms. (Top) Actual experimental sequence, with wait times inserted for camera readout. (Bottom) Projected sequence duration, totaling less than \TimeProjectedThreeOutcomeMeasurement.}
\end{figure*}

\begin{table*}
\caption{\label{table1}%
\textbf{Detection probability of the three outcome measurement.}
This table provides the numerical values corresponding to the data shown in Fig.~\ref{fig2}b. While preparation erasure detection is applied, we do not correct for spin preparation errors arising from the $\pi$-pulse in the metastable state manifold used to prepare $\ket{m_0}$.}
\begin{tabular}{|c|| c| c|c|} 
\hline
 \textbf{Prepared} & \textbf{\mone observed} & \textbf{\mzero observed} & \textbf{Loss observed }\\ [0.5ex] 
 \hline
 \mone  & 96.9(2) \% & 0.18(6) \%&2.8(2) \% \\ 
 \hline
 \mzero & 0.70(10)\% & 94.9(2) \% & 4.3(2) \% \\
 \hline
 Non  & 0.02(3)\% & 0.02(3)\% & 99.97(3)\% \\
 \hline
\end{tabular}
\end{table*}

\begin{table*}
\caption{\label{table2}%
\textbf{Preparation error summary.}
Errors from individual preparation steps, extracted from experimental measurements and/or calculations. Experimentally characterized (calculated) errors are indicated by [e] ([c]). The listed errors can largely be separated into those that are detectable via erasure and those that are not.  The third column indicates spin-flip errors that occur during a state preparation step. The last two rows show the additional preparation error that occurs if one prepares \mzero. }
\begin{tabular}{|c|| c| c|c|} 
\hline
 \textbf{Preparation step} & \textbf{Loss Error} & \textbf{Erasure corrected} & \textbf{Spin flip error}\\ [0.5ex] 
 \hline
 Imaging loss [e]  & 0.4(1) \% & 0.4(1) \% & - \\ 
 \hline
  $g$-qubit $\pi$-pulse [e] & 0.9(1)\% & $<$0.02\% & - \\
 \hline
 Tweezer ramp down [e] & 0.4(1)\% & 0.4(1)\%& -  \\
 \hline
 $o$-qubit MPP ($\sigma^+$-pol)  [e] & 0.4(1)\% & $<$0.01\%& - \\
 \hline
 $o$-qubit  MPP ($\pi$-pol.) [e] & -  & - & 0.19(6)\%\\
 \hline
 \textbf{Total \mone} & \textbf{2.1(2)\%} & \textbf{0.8(1)\%} &  \textbf{0.19(6)\%}\\
 \hline
 $m$-qubit $\pi$-pulse [e+c] & 0.56(1)\%  & 0.38(1)\% & 0.38(1)\%\\
 \hline
\textbf{Total \mzero } & \textbf{2.7(2)\%} & \textbf{1.2(1)\%} &  \textbf{0.57(6)\%}\\
 \hline
\end{tabular}
\end{table*}

\begin{table*}
\caption{\label{table3}%
\textbf{Readout error summary.}
Errors from individual readout steps, extracted from separate measurements. Experimentally characterized (calculated) errors are indicated by [e] ([c]). Each error contributes either to atom loss (``Loss error") or to spin mislabeling (``Spin misl. error"), where spin mislabeling refers to the incorrect assignment of a spin state. The final two columns indicate how each error source affects the first and second spin readouts, respectively.}
\begin{tabular}{|c|| c| c|c|c|} 
\hline
 \textbf{Readout step} &  \textbf{Loss Error} &  \textbf{Spin misl. Error}  & \textbf{Readout \mone} & \textbf{Readout \mzero}  \\ [0.5ex] 
 \hline
 Release and Recapture [e]  & 0.3(1) \% & -  &   1x &  2x \\ 
 \hline
 $o$-qubit $\pi$-pulse ($\pi$-pol) [e/c] & x & 0.10(3) \% &  0x &  1x \\
 \hline
  $o$-qubit $\pi$-pulse ($\sigma^+$-pol) [c] & 0.4(1)\% &  - & 1x &  1x \\
 \hline
  $m$-qubit $\pi$-pulse [e] & 0.56(1)\% & - &  0x &  1x \\
 \hline
 Destructive image infidelity [e] & 0.2(1)\% & -  & 1x  &  1x\\
 \hline
  Raman scattering deep tweezer [c] (\ground + \clock) & 0.13(1)\% & - & 0x &  1x \\
 \hline
  Raman scattering hold (per ms) [c] (\ground + \clock) & 0.005(1)\% & - & 5x &  13x \\
 \hline
  Vacuum loss (per ms)[c] & 0.00026(1)  & - & 5x  &  13x\\
 \hline
 \textbf{Total readout loss/ Spin misl. [\%]} & &    & \textbf{0.9(2)/0.0} &\textbf{1.9(2)/0.10(3)} \\
 \hline
 Preparation Error [\%] &  &   & 0.8(1)/0.19(6)  &  1.2(1)/0.57(6) \\
 \hline
 \textbf{Total loss/ Spin misl. [\%]} & &    & \textbf{1.7(2)/0.19(6)} &\textbf{3.1(2)/0.7(1)} \\
 \hline
\end{tabular}
\end{table*}

\begin{figure*}
    \centering
    \includegraphics[width=\columnwidth]{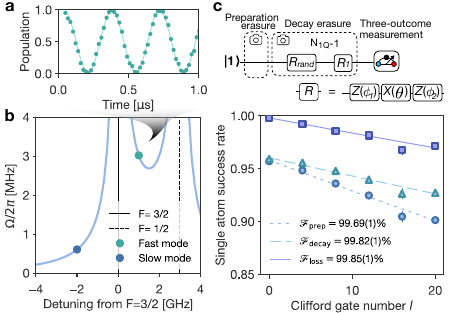}
    \caption{\label{SMfig3}%
\textbf{Single-qubit control.}
\textbf{a,}~Fast Rabi oscillations on the metastable-state qubit, as used for mapping from the Rydberg state. The fit yields a Rabi frequency of \FreqMetaRabiFast{}.
\textbf{b,}~Calculated dependence of the Rabi frequency on detuning, at constant laser power. When high speed is not required—such as in global randomized benchmarking (gRB) of the two-qubit gate—we use a slower Rabi frequency by increasing the detuning, in order to reduce intermediate-state scattering and mitigate the finite AOM turn-on time.
\textbf{c,}~Single-qubit Clifford randomized benchmarking comparing erasure and three-outcome measurement schemes. To reduce scattering errors, $Z$-rotations are implemented via phase tracking.}
\end{figure*}

\begin{figure*}
    \centering
    \includegraphics[width=\textwidth]{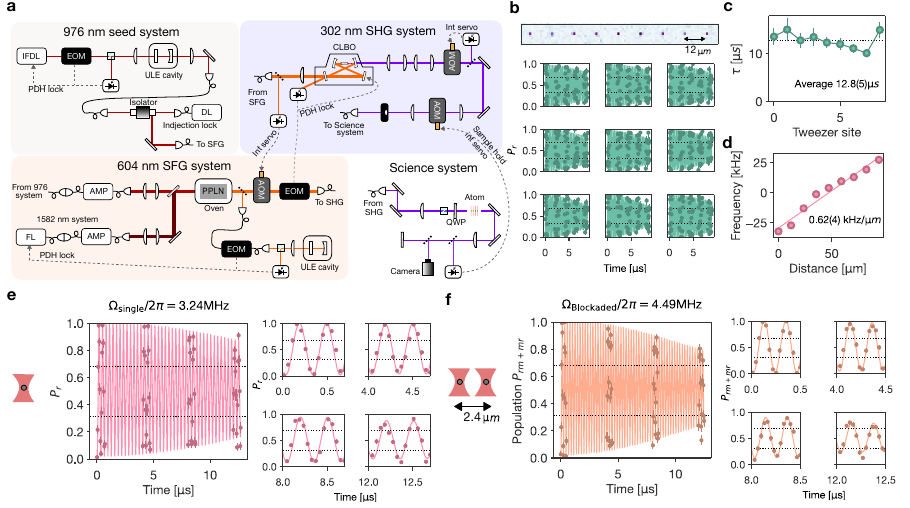}
    \caption{\label{SMfig4}\textbf{Rydberg laser system and coherent manipulation of the Rydberg state.}
\textbf{a,}UV laser system. We use fiber lasers (FL) and diode lasers (DL) as seed sources, which are amplified using fiber amplifiers (AMP). High-power beams are then combined via sum frequency generation (SFG) and frequency-doubled using second harmonic generation (SHG) to produce several hundred milliwatts of UV light. To suppress frequency noise, the cavity transmission of a high-finesse ($\mathcal{F} = \NumbersFinesseULERydberg$) ultra-low-expansion (ULE) cavity is used to seed the \WavelengthUVFundamentalOne system~\cite{levine2018high}. A PPLN crystal is used for SFG of the \WavelengthUVFundamentalOne and \WavelengthUVFundamentalTwo beams, generating \WavelengthUVSFG light, which is then frequency-doubled in an SHG cavity containing a CLBO crystal. To extend the crystal lifetime, the SHG cavity is unlocked between experiments, stopping UV generation. Electro-optic modulators (EOMs) are used for all laser frequency locks. UV intensity stabilization is achieved via servo control of both the \WavelengthUVSFG{} and final \WavelengthUV beams. A sample-and-hold feedback scheme is used to suppress shot-to-shot pulse fluctuations.
\textbf{b,}~$T_2^*$ coherence assessment of the Rydberg transition. Atoms are spaced for about \LengthscaleRamsey, and a Ramsey sequence is performed with a detuning offset of \FreqRamseyDark applied during the dark time.
\textbf{c,}~Observed coherence time across the tweezer array, with an average of \TimeRamsey.
\textbf{d,}~Variation of the oscillation frequency across the array reveals an energy gradient of \NumbersEFieldGradient.
\textbf{e,}~Single-atom Rydberg Rabi oscillations on the $r$-qubit under loss detection. We observe \NumbersRabiSingleFlip coherent oscillations before $1/e$ decay.
\textbf{f,}~Rydberg Rabi oscillations under blockade. Due to interaction, the Rabi frequency is enhanced by a factor of $\sqrt{2}$ compared to the non-interacting case. We observe \NumbersRabiBlockadedFlip coherent oscillations before $1/e$ decay.}
\end{figure*}

\begin{figure*}
    \centering
    \includegraphics[width=\columnwidth]{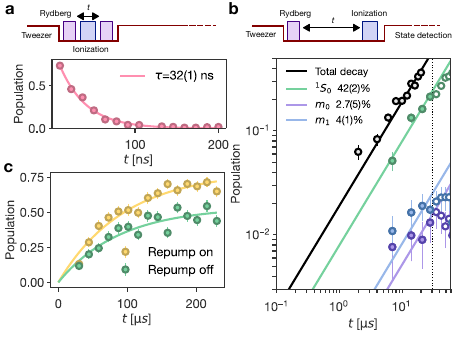}
    \caption{\label{SMfig4_5}\textbf{Auto-ionization and Rydberg state decay branching. }\textbf{a,} Characterization of the population decay, varying the ionization pulse duration in between the $\pi$-pulses for the Rydberg transition. The solid line is an exponential fit. \textbf{b,}~Characterization of the branching of the decayed Rydberg state to ground state and metastable state. Black lines are total amount of decay measured by two consecutive Rydberg $\pi$-pulses. Remaining branching is measured by the sequence shown on top of the graphic, where we changed the timing of the ionization of the Rydberg state and measured the amount of the remaining population. \textbf{c,} Assessment of the branching to the $^3P_2~F=3/2$ state. Decay experiments by monitoring the ground state. In one case we apply a 770 nm repump laser resonant with the $^3\mathrm{P}_2~F=3/2\rightarrow{}^3\mathrm{S}_1$ transition, and in another case we do not apply the repumper. The ratio of the two experiments allow us to estimate the decay population to $^3\mathrm{P}_2~F=3/2$.
}
\end{figure*}

\begin{figure*}
    \centering
    \includegraphics[width=\columnwidth]{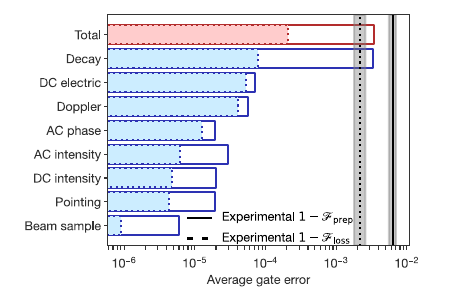}
\caption{\label{SMfig5}%
\textbf{Two-qubit gate error assessment.} Theoretical assessment of known error contributions. Simulated gate infidelities are shown both without loss detection (solid bars) and with loss detection (dotted bars). The top row presents the average gate fidelity from full error-model simulations. Individual error sources are ordered by their contribution to the loss-detected error. We also show the experimental result from Fig.~\ref{fig3} with vertical lines.}
\end{figure*}

\begin{figure*}
    \centering
    \includegraphics[width=\columnwidth]{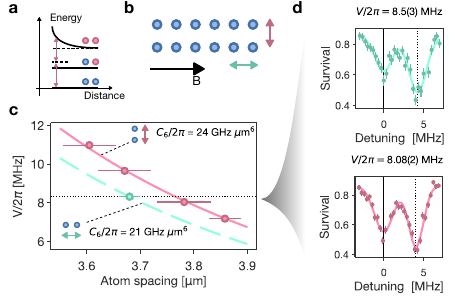}
    \caption{\label{SMfig6}%
    \textbf{Rydberg interaction calibration.}
    \textbf{a,}~ For the interaction calibration, we use a two-photon transition to the doubly excited state.  \textbf{b,} As reported on~\cite{peper2025spectroscopy}, we observe an anisotropy of the interaction depending on the orientation of the interaction axis relative to the magnetic field. Adjusting the atom distances effectively generates the square lattice interaction. \textbf{c,} Result of the anisotropy calibration. Showing higher interaction strength for the direction perpendicular to the magnetic field. \textbf{d,} The interaction spectroscopy result after calibrating the lattice geometry. Residual imbalance is caused by the limited SLM discretization, which can be solved by using more computational resources. }
\end{figure*}

\begin{figure*}
    \centering
    \includegraphics[width=\textwidth]{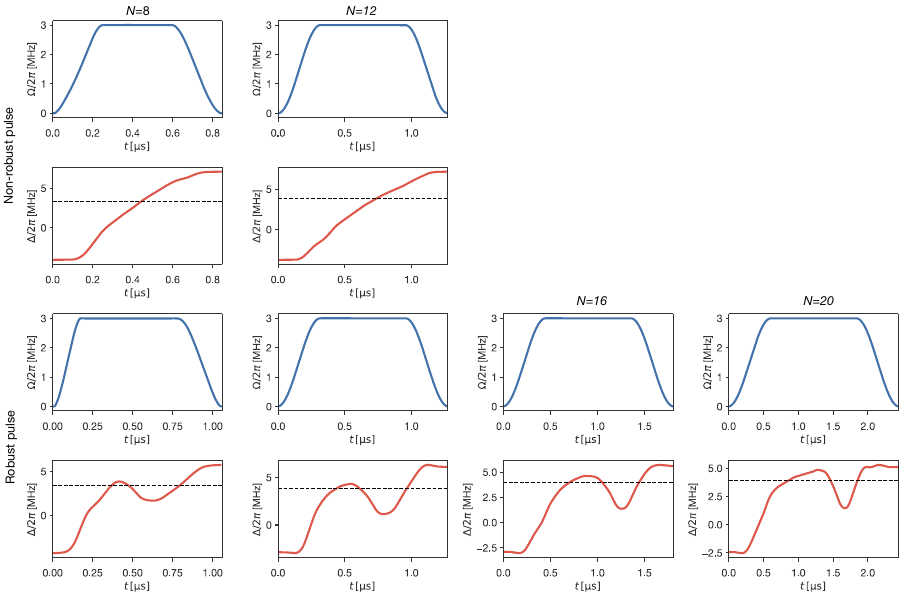}
    \caption{\label{SMfig7_0}%
    \textbf{Implemented optimal control pulse.}
    Rabi frequency (top) and detuning (bottom) sweep profiles are shown for both robust and non-robust pulses implemented in Fig.~\ref{fig4}. The dotted line is a theoretically estimated phase transition point.}
\end{figure*}

\begin{figure*}
    \centering
    \includegraphics[width=\textwidth]{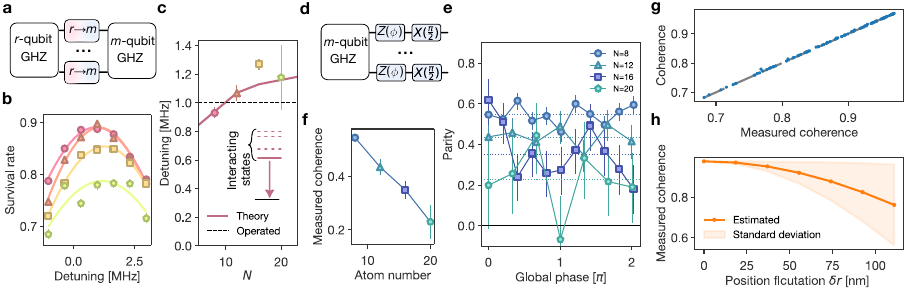}
    \caption{\label{SMfig7}%
    \textbf{Qubit mapping of the $Z_2$-GHZ state and assessment of the coherence.}
    \textbf{a,}~ Schematic of the qubit mapping of the $Z_2$-GHZ state. \textbf{b,}~ To achieve maximum mapping efficiency a detuned $\pi$-pulse is used for the mapping. \textbf{c,}~ Summary of the relation between atom number and optimal detuning. We choose \NumberProjectionDetuning for all of the experiments. The inset shows the schematic of the qubit mapping. \textbf{d,}~ Schematic of the sequence for the assessment of the GHZ coherence. After mapping to the non-interacting metastable state qubit, we apply a global $\pi/2$-pulse with various phases. \textbf{e,}~ Result of the parity measurements for the GHZ coherence assessment. We estimate the coherence of the GHZ state by averaging the observed parity expectation values after a global phase shift, over the interval 0 to $2\pi$. \textbf{g,}~ Theoretical comparison of actual coherence and estimated coherence by the presented method. \textbf{h,}~ GHZ coherence for various distance fluctuations. $\delta r$ is the standard deviation of the Gaussian distribution of the shot-to-shot position fluctuations. }
\end{figure*}

\begin{figure*}
    \centering
    \includegraphics[width=\textwidth]{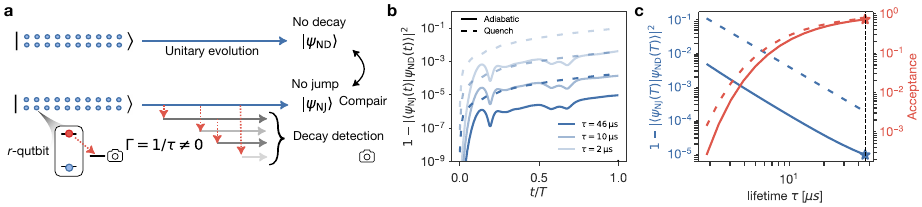}
\caption{\label{SMfig8}%
\textbf{Analysis of the effect of decay detection on the many-body preparation dynamics.} \textbf{a,}~ Concept of the simulation. We theoretically compare the many-body evolution in the case of no Rydberg decay (ND) and the case of finite Rydberg decay and perfect decay detection. The latter dynamics is equivalent to the evolution generated by an effective non-Hermitian Hamiltonian describing the no-jump (NJ) component of the dissipative process.
We vary the magnitude of the decay rate and examine how the prepared states deviate from the no-decay case by computing the infidelity $1-|\langle \psi_\mathrm{NJ} | \psi_\mathrm{ND} \rangle|^2$. \textbf{b,}~ Infidelity during the evolution for various decay rates. For the adiabatic evolution (solid lines), we analyze the many-body evolution generating a $20$-atom GHZ state with the robust preparation protocol for $\Omega T = 46$. This is what is used for the $N=20$ GHZ state in Fig.~\ref{fig4}. For the quench evolution (dashed lines), we set zero Rydberg detuning and constant Rabi frequency at $2\pi\times3$~\MHz. \textbf{c,}~ Infidelity at the end of the preparation. The vertical dashed line ($\tau = 46 \mu \mathrm{s}$) denotes the Rydberg state lifetime in the present experiment. Stars indicate the condition implemented in the GHZ generation experiment.}
\end{figure*}

\end{document}